\documentclass[11pt, oneside]{article}   	
\usepackage{geometry} 
\usepackage{amsmath} 								
\usepackage{graphicx}	
\usepackage{amssymb}
\usepackage{cite}
\usepackage{color,xspace}

\numberwithin{equation}{section}
\DeclareSymbolFont{extraup}{U}{zavm}{m}{n}
\DeclareMathSymbol{\vardiamond}{\mathalpha}{extraup}{87}
\usepackage{float}
\restylefloat{table}
\allowdisplaybreaks

\def\twomat[#1,#2][#3,#4]{\left( \begin{array}{cc} #1 & #2 \\ #3 & #4 \end{array} \right)}
\def\thv[#1,#2,#3]{\left( \begin{array}{c} #1 \\ #2 \\ #3 \end{array} \right)}
\def\twv[#1,#2]{\left( \begin{array}{c} #1 \\ #2 \end{array} \right)}
\def\lagr{\mathcal{L}}
\def\nn{\nonumber}
\def\ov{\overline}
\newcommand{\SARAH}{{\tt SARAH}\xspace}
\def\DR{\ensuremath{\overline{\mathrm{DR}}}\xspace}
\def\MS{\ensuremath{\overline{\mathrm{MS}}}\xspace}
\def\MSUSY{\ensuremath{M_{\rm SUSY}}\xspace}
\def\blog{\overline{\log}}
\def\GeV{\ensuremath{\mathrm{GeV}}}
\def\TeV{\ensuremath{\mathrm{TeV}}}

\title{Higgs alignment from extended supersymmetry}

\date{}

\begin{document}

\begin{flushright}
\end{flushright}
\begin{center}

\vspace{1cm}
{\LARGE{\bf Higgs alignment from extended supersymmetry}}

\vspace{1cm}

\large{Karim Benakli$^\spadesuit$ \let\thefootnote\relax\footnote{$^\spadesuit$kbenakli@lpthe.jussieu.fr},
Mark D. Goodsell$^{\vardiamond}$ \let\thefootnote\relax\footnote{$^\vardiamond$goodsell@lpthe.jussieu.fr}
and Sophie~L.~Williamson$^\clubsuit$ \footnote{$^\clubsuit$swilliamson@lpthe.jussieu.fr}
 \\[5mm]}

{ \sl Laboratoire de Physique Th\'eorique et Hautes Energies (LPTHE),\\ UMR 7589,
Sorbonne Universit\'e et CNRS, 4 place Jussieu, 75252 Paris Cedex 05, France.}

\end{center}
\vspace{0.7cm}

\abstract{We consider the effective type-II Two-Higgs doublet model originating from Dirac gaugino models with extended supersymmetry in the gauge sector, which is automatically aligned in the simplest realisations. We show that raising the scale at which the extended supersymmetry is manifest and including quantum corrections actually improves the alignment. Using an effective field theory approach including new threshold corrections and two-loop RGEs, plus two-loop corrections to the Higgs mass in the low-energy theory, we study the implications from the Higgs mass and other experimental constraints on the scale of superpartners. We contrast the results of the minimal Dirac gaugino model, where alignment is automatic, with the hMSSM and the MRSSM, where it is not, also providing an hMSSM-inspired analysis for the new models.}


\newpage
\setcounter{footnote}{0}

\section{Introduction}
\label{introduction}


In the absence of signals of strongly-coupled particles at the LHC, it has become important to study the possibility of new particles that couple to Standard Model (SM) states only via couplings of electroweak strength. The bounds on such particles are still relatively weak but with much luminosity to arrive there is still a substantial parameter space to explore, and such theories perhaps represent now the best chance for discoveries. Among such theories, one that has received significant and now increasing attention is the Two Higgs Doublet Model (THDM); see e.g. \cite{Gunion:1989we,Davidson:2005cw,Bernon:2015qea,Bernon:2015wef} and references therein. It is important to ask the question: ``does the Higgs sector just consist of one doublet?'' because the answer will give profound information about nature. If there are indeed additional fundamental scalars that mix with the Higgs boson, then this dramatically worsens the Hierarchy problem and would necessitate a rethinking of our ideas of naturalness. On the other hand, such sectors naturally appear in the context of supersymmetry (SUSY) and it is conceivable that a second Higgs doublet could be the harbinger of a full SUSY theory.

However, the measurements of the Higgs boson's \emph{couplings} already place significant constraints on the amount of mixing that it can suffer. It is for this reason that there has been much interest in the idea of \emph{alignment} in the Higgs sector, i.e. that the mass eigenstates align with the vacuum expectation value, because in this case the couplings would be exactly SM-like. 

To quantify this, consider two Higgs doublets $\Phi_1, \Phi_2$ which mix, and then rotate their neutral components as follows:
\begin{align}
 \twv[\mathrm{Re}(\Phi_1^0),\mathrm{Re}(\Phi_2^0)] = \frac{1}{\sqrt{2}}\twomat[c_\beta,-s_\beta][s_\beta,c_\beta]\twv[v+\tilde{h},\tilde{H}]
\end{align}
where we shall throughout use the notation 
$$ c_\beta \equiv \cos \beta, \qquad s_\beta \equiv \sin \beta, \qquad t_\beta \equiv \tan \beta.$$
In this basis, we can write the mass matrix as 
\begin{eqnarray}
\mathcal{M}^2_h \equiv \begin{pmatrix}
Z_1 v^2 & Z_6 v^2 \\
Z_6 v^2 & m_A^2 + Z_5 v^2 \end{pmatrix}, 
\end{eqnarray}
where the quantities $Z_1, Z_5, Z_6$ are functions of the quartic couplings and mixing angles only; we shall give explicit expressions for this relationship later, in equation (\ref{EQ:THDMZis}).
Clearly the mass eigenstates are only $\tilde{h},\tilde{H}$ if 
$$ Z_6 = 0,$$
and this is the condition for alignment, because the fields \emph{align} with the electroweak vacuum expectation value. On the other hand, if $Z_6 \ne 0$, we must make a further rotation which is conventionally parameterised by an angle $\alpha$ as
\begin{align}
\twv[\tilde{h},\tilde{H}] =& \twomat[ s_{\beta - \alpha}, c_{\beta - \alpha}][c_{\beta - \alpha},-s_{\beta - \alpha}] \twv[h,H] 
\end{align}
where now $h,H$ are the two mass eigenstates. We shall assume throughout that $h$ is the lightest eigenstate. In terms of the masses of the physical bosons $m_{h,H}$ this gives 
\begin{align}
Z_6 v^2 =& s_{\beta - \alpha} c_{\beta - \alpha} (m_{h}^2 - m_{H}^2).
\end{align}
In both the type-I and type-II THDM, there is a Higgs eigenstate that couples to the up-type quarks, and we define this eigenstate to be $\Phi_2$. This means that the ratio of the  $h$ coupling to all up-type quarks compared to the  SM Higgs' value is 
$$
\kappa_u = \frac{\cos \alpha}{\sin \beta},
$$
while the ratio of the coupling to vector bosons to the SM value is also determined entirely by the mixing (neglecting loop effects from the rest of the extended Higgs sector):
\begin{align}
\kappa_V = \sin (\beta - \alpha).
\end{align}
However, there is a combined ATLAS+CMS bound \cite{Khachatryan:2016vau} on the ratio of these:
\begin{align}
\lambda_{Vu} \equiv \frac{\kappa_V}{\kappa_u} = 1^{+0.13}_{-0.12} = \frac{1}{1+ \frac{1}{t_\beta t_{\beta - \alpha}}}.
\end{align}
This is enough to constrain
\begin{align}
t_\beta t_{\beta - \alpha} \gtrsim 7.3 \Rightarrow |Z_6| \lesssim \bigg| - \frac{7.3 t_\beta}{53 + t_\beta^2}  \frac{m_H^2 - m_h^2}{v^2} \bigg| \lesssim  \bigg|-0.5 \frac{m_H^2 - m_h^2}{v^2} \bigg|,
\end{align}
where the latter bound comes from the value $t_\beta = 7.3$, and the bound is much more stringent for large or small $t_\beta$. 
For $m_H$ somewhat above $m_h$ this is a rather weak constraint, only becoming relevant when the two states approach degeneracy. However, in the type-II THDM, there is another constraint from the ratio of the ratio of the neutral Higgs coupling to all down-type quarks compared to its SM value
$$ \kappa_d = - \frac{\sin \alpha}{\cos \beta}$$
via
\begin{align}
\lambda_{du} \equiv \frac{\kappa_d}{\kappa_u} = 0.92 \pm 0.12 = \frac{1 - \frac{t_\beta}{t_{\beta - \alpha}}}{1 + \frac{1}{t_\beta t_{\beta - \alpha}}}
\end{align}
and, since from the previous constraint we know that the denominator is nearly equal to one, we have
\begin{align}
-0.04 \lesssim \frac{t_\beta}{t_{\beta - \alpha}} \lesssim 0.2
\end{align}
which in turn implies $t_{\beta - \alpha} \gg t_\beta$ and so $s_{\beta - \alpha} c_{\beta - \alpha} \simeq \frac{1}{t_{\beta - \alpha}}$ and 
\begin{align}
-0.04 \frac{m_H^2 - m_h^2}{t_\beta v^2}  \lesssim Z_6 \lesssim 0.2 \frac{m_H^2 - m_h^2}{t_\beta v^2} .
\end{align}
This leads to a sensible constraint; for example, for $m_H = 600$ GeV and $t_\beta =5$ it leads to $Z_6 \lesssim 0.2$. So we see that either we should take the mass $m_H$ to be large, in which case we have \emph{decoupling}, or we keep it light in order to possibly detect it at the LHC, in which case we need \emph{alignment without decoupling}  (see e.g. \cite{Carena:2013ooa}). However, as we have seen this is non-trivial; as the LHC measurements become more precise, the constraints will tighten further, and it is in this spirit that it is important to consider models where the alignment is \emph{natural} rather than \emph{ad hoc}.


The problem for the different types of THDM is that alignment without decoupling is not generic when we choose the masses -- or equivalently quartic couplings -- from the bottom up. Hence it is logical to derive the couplings of the THDM from some higher-energy theory and look for cases where alignment arises naturally.  For example, \cite{Dev:2014yca,Pilaftsis:2016erj,Das:2017zrm} proposed models which lead to a natural alignment condition, based on additional bosonic symmetries. Here, on the other hand, we shall show how alignment arises \emph{automatically} in a class of supersymmetric models, in contrast to the MSSM or NMSSM \cite{Carena:2015moc}, with the additional benefits of (greatly) increasing the naturalness of the model and being able to predict the scale of new superpartners. Moreover, we shall show that quantum corrections actually \emph{improve} the alignment!

The class of models that we shall consider have a gauge sector which is enhanced to $N=2$ supersymmetry at a (potentially high) scale $M_{N=2}$. This fits into the framework of Dirac gaugino models, which have been well-studied in, for example, \cite{Fayet:1978qc,Polchinski:1982an,Hall:1990hq,Fox:2002bu,Nelson:2002ca,Antoniadis:2006uj,Kribs:2007ac,%
Amigo:2008rc,Plehn:2008ae,Benakli:2008pg,Belanger:2009wf,Benakli:2009mk,Choi:2009ue,Kribs:2010md,Fok:2010vk,Benakli:2010gi,Choi:2010gc,%
Carpenter:2010as,Abel:2011dc,Davies:2011mp,Benakli:2011vb,Benakli:2011kz,Kalinowski:2011zz,Frugiuele:2011mh,%
Itoyama:2011zi,Rehermann:2011ax,Bertuzzo:2012su,Davies:2012vu,Argurio:2012cd,Fok:2012fb,Argurio:2012bi,Frugiuele:2012pe,%
Frugiuele:2012kp,Benakli:2012cy,Dudas:2013gga,Arvanitaki:2013yja,Itoyama:2013sn,Chakraborty:2013gea,Csaki:2013fla,Itoyama:2013vxa,Beauchesne:2014pra,%
Benakli:2014daa,%
Bertuzzo:2014bwa,Diessner:2014ksa,Benakli:2014cia,Goodsell:2014dia,Busbridge:2014sha,Chakraborty:2014sda,Ding:2015wma,Alves:2015kia,Alves:2015bba,Carpenter:2015mna,Martin:2015eca,Goodsell:2015ura,
Diessner:2015yna,Diessner:2015iln,Kotlarski:2016zhv,Benakli:2016ybe,Braathen:2016mmb,Diessner:2017ske,Chakraborty:2017dfg}. In particular, the idea of $N=2$ supersymmetry in the gauge sector only and the consequences for the Higgs sector were first explored in \cite{Antoniadis:2006uj} and recently studied in \cite{Heikinheimo:2011fk,Ellis:2016gxa}. In general, though, this was either taken to be at the same scale as the other superpartners \cite{Ellis:2016gxa}, or only a rough estimate of the main contribution of the chiral sector was included \cite{Antoniadis:2006uj}, while we shall show that increasing $M_{N=2}$ improves alignment and increases naturalness! 

In section \ref{diracgauginos_secInteractions} we will describe our theory and how it leads to natural alignment at tree level. In section \ref{SEC:RADIATIVE} we will outline the effect of radiative corrections. In section \ref{SEC:PRECISION} we perform a precision study of the model using an EFT approach to obtain the parameters at low energies, give predictions for the scale of new physics from the value of the Higgs mass, and explore the consequences for alignment. In section \ref{SEC:CONSTRAINTS} we consider all of the relevant constraints on the model space, including the latest LHC search for decays to $\tau$ pairs, $b\rightarrow s \gamma$ searches and electroweak precision constraints, and show how this affects our model. In the appendix we give all of the one-loop threshold corrections for our model at the scale of supersymmetry. Finally, in section \ref{SEC:MRSSM} we briefly consider the case of the MRSSM.

\section{Alignment from extended supersymmety}
\label{diracgauginos_secInteractions}

\subsection{The Higgs sector of Dirac gaugino models}

\subsubsection{The minimal model}

To endow gauginos with a Dirac mass, at a minimum we need to add chiral fermions in the adjoint representation of each gauge group, which means adding adjoint chiral superfields: a singlet $\mathbf{S}$, an $SU(2)$ triplet  $\mathbf{T}$, and an $SU(3)$ octet $\mathbf{O}$. If we add just these fields, then we have the simplest Dirac-gaugino extension of the MSSM whose Higgs sector has been well studied \cite{Belanger:2009wf,Benakli:2010gi,Benakli:2011kz,Benakli:2012cy,Braathen:2016mmb}. However, we can then choose the superpotential according to the symmetries that we want to preserve. One motivation for the adjoint fields is as the additional degrees of freedom from an $N=2$ supersymmetric gauge multiplet, and then the $H_u, H_d$ fields become an $N=2$ hypermultiplet; in this work we shall assume that $N=2$ supersymmetry \emph{in the gauge/Higgs sector only} is valid above some scale $M_{N=2}$. In this case, we can immediately write down the superpotential
\begin{eqnarray}
W_{\text{Higgs}} =  \mu \,  \bold{H_u} \cdot \bold{H_d}+ \lambda_S \bold{S} \, \bold{H_u} \cdot \bold{H_d} + 2 \lambda_T \, \bold{H_d} \cdot \bold{T} \bold{H_u} \label{W_Higgs}
\end{eqnarray} 
which contains the only interactions compatible with $N=2$ SUSY and includes a central role for the R-symmetry. Indeed, under the R-symmetry of the $N=1$ theory the adjoint scalars must have zero charge, and this prevents couplings of the form $S^2, S^3$ etc which would otherwise be permitted by the gauge symmetry. 
The condition of $N=2$ supersymmetry imposes 
\begin{align}
\lambda_S= \frac{1}{\sqrt{2}} g_Y, \qquad \lambda_T =\frac{1}{\sqrt{2}} g_2
\end{align}
(where $g_Y, g_2$ are the hypercharge and $SU(2)$ gauge couplings) \emph{at the scale $M_{N=2}$}, which we shall in general take to be greater than the $N=1$ SUSY scale. 

We must also add supersymmetry-breaking terms, and these do not necessarily need to respect the same symmetries as supersymmetric terms. The most general choice that we can make for the Higgs and adjoint scalar sector for the \emph{standard} soft terms is
\begin{align}
&\lagr_{\rm standard\ soft} = m_{H_u}^2 |H_u|^2 + m_{H_d}^2 |H_d|^2 + B_{\mu} (H_u \cdot H_d + \text{h.c}) + \frac{1}{2} M_i \lambda_i \lambda_i \\ 
 & + m_S^2 |S|^2 + 2 m_T^2 \text{tr} (T^{\dagger} T) + \frac{1}{2} B_S \left(S^2 + h.c\right)+  B_T\left(\text{tr}(T T) + h.c.\right)  + m_O^2 |O|^2 + B_O\left(\text{tr}(O O) + h.c.\right)\nonumber\\
& + A_S \left(S H_u \cdot H_d + h.c \right) + 2 A_T  \left( H_d \cdot T H_u + h.c \right) + \frac{A_\kappa}{3} \left( S^3 + h.c. \right) + A_{ST} \left(S \mathrm{tr} (TT) + h.c \right) +  A_{SO} \left(S \mathrm{tr} (OO) + h.c \right), \nn
\end{align} 
where $\lambda_i = \{\lambda_Y, \lambda_2, \lambda_3\}$ are the gauginos of hypercharge, $SU(2)$ and $SU(3)$ respectively, with Majorana masses $M_{Y}, M_2, M_3$, and to these we add the \emph{supersoft} operators $m_{Di }\theta^\alpha$ for Dirac masses as
\begin{align}
\int d^2\theta \left[  \sqrt{2} \, m_{DY} \theta^\alpha \bold{W}_{1\alpha} \bold{S} + 2 \sqrt{2} \, m_{D2}\theta^\alpha \text{tr} \left( \bold{W}_{2\alpha} \bold{T}\right)  +  2 \sqrt{2} \, m_{D3}\theta^\alpha \text{tr} \left( \bold{W}_{3\alpha} \bold{O}\right)\! \right]
\end{align} 
where $\bold{W}_{i\alpha}$ are the supersymmetric gauge field strengths.

Since we are interested in \emph{Dirac} gaugino masses and their attractive theoretical and phenomenological properties, we should expect that the terms that violate R-symmetry should be small: this includes the Majorana gaugino masses; $A_S, A_T$; but also $B_\mu$. However, we require that the R-symmetry \emph{is} broken at some scale, since we believe that global symmetries cannot be exact; but also, in this model, the Higgs must carry R-charge and so the absence of an R-axion requires it. Indeed, the R-axion is essentially the Higgs pseudoscalar, whose mass is controlled by the $B_\mu$ term. We therefore, as in earlier works, take $B_\mu$ to have a small but non-zero value. We can also take motivation from models of gauge mediation of supersymmetry \cite{Benakli:2008pg,Benakli:2016ybe}, where the trilinears are all small, and we shall mostly neglect them in the following (although they do not significantly affect the analysis). 

On the other hand, in gauge-mediated models the adjoint scalars are typically the heaviest states. Taking large $m_{S}, m_T, m_O$ then motivates integrating them out of the light spectrum. Interestingly, since $B_\mu$ should remain small due to the approximate R-symmetry, if we were to tune the Higgs masses such that only one remains light, then we would have very large $\tan \beta$, and would have trouble obtaining the correct Yukawa couplings for the down-type quarks and leptons. This implies that a second Higgs should be taken to be somewhat light, and motivates studying the two-Higgs doublet limit of the model. 

Finally, we note that this model does not have gauge-coupling unification. If we wish to naturally restore gauge coupling unification, we can add additional vector-like lepton fields, as was done in \cite{Benakli:2014cia,Goodsell:2015ura}. Since they are vector-like, we could also allow them to be hypermultiplets of the $N=2$ at $M_{N=2}$, but their inclusion will little change the discussion in this paper so for sake of generality we shall neglect them.

\subsubsection{The MRSSM}

Another very popular realisation of Dirac gaugino models is the MRSSM \cite{Kribs:2007ac,Diessner:2014ksa,Diessner:2015yna,Diessner:2015iln,Diessner:2017ske}. In this model, we preserve an exact continuous R-symmetry by including some R-Higgs doublet superfields which couple to the Higgs bosons but do not obtain an expectation value, allowing the Higgs doublets $H_u, H_d$ to have zero R-charge. The Higgs superpotential becomes\footnote{We note the discrepancy of a factor of $2$ for the triplet coupling terms compared to  \cite{Diessner:2014ksa, Bertuzzo:2014bwa}, which arises due to a difference in definition of $T$ and the choice for the neutral components to take the same pre-factor as the singlet neutral components.} 
\begin{eqnarray}
W_{\text{Higgs}}^{\rm MRSSM} &=&  \mu_u \,  \bold{R_u} \cdot \bold{H_u} +  \mu_d \,  \bold{R_d} \cdot \bold{H_d} + \lambda_{S_u} \bold{S} \, \bold{R_u} \cdot \bold{H_u} + \lambda_{S_d} \bold{S} \, \bold{R_d} \cdot \bold{H_d} \nonumber \\
& & + 2 \lambda_{T_u} \, \bold{R_u} \cdot \bold{T} \bold{H_u} + 2 \lambda_{T_d} \, \bold{R_d} \cdot \bold{T} \bold{H_d}\,.  \label{W_Higgs_MRSSM}
\end{eqnarray}
If we then impose $N=2$ supersymmetry at some scale, we can treat $(R_u, H_u)$ and $(R_d, H_d)$ as hypermultiplets and then we would have
\begin{align}
\lambda_{S_u} = \frac{g_Y}{\sqrt{2}}, \qquad  \lambda_{S_d} = -\frac{g_Y}{\sqrt{2}}, \qquad \lambda_{T_u} =  \lambda_{T_d} = \frac{g_2}{\sqrt{2}}.,
\end{align}
where the difference in sign is explained by the different charges of the hypermultiplets.\footnote{Equivalently the second hypermultiplet could be written $(H_d, R_d)$ and then we would have $\lambda_{S_d} = g_Y/\sqrt{2}, \lambda_{T_d} = -g_2/\sqrt{2}.$}

R-symmetry then limits the possible soft-supersymmetry breaking terms to consist of only the supersoft operator, squark/slepton masses and 
\begin{align}
&\lagr_{\rm standard\ soft}^{\rm MRSSM} = m_{H_u}^2 |H_u|^2 + m_{H_d}^2 |H_d|^2 + B_{\mu} (H_u \cdot H_d + \text{h.c}) + m_{R_u}^2 |R_u|^2 + m_{R_d}^2 |R_d|^2\\ 
 & + m_S^2 |S|^2 + 2 m_T^2 \text{tr} (T^{\dagger} T) + \frac{1}{2} B_S \left(S^2 + h.c\right)+  B_T\left(\text{tr}(T T) + h.c.\right)  + m_O^2 |O|^2 + B_O\left(\text{tr}(O O) + h.c.\right)\nonumber\\
& + A_S \left(S H_u \cdot H_d + h.c \right) + 2 A_T \left( H_d \cdot T H_u + h.c \right) + \frac{A_\kappa}{3} \left( S^3 + h.c. \right) + A_{ST} \left(S \mathrm{tr} (TT) + h.c \right) +  A_{SO} \left(S \mathrm{tr} (OO) + h.c \right). \nn
\end{align} 
The terms on the last line are usually neglected, but there is no symmetry that forbids them (even if we expect them to be small in e.g. gauge mediation models).


\subsection{Two-Higgs doublet model limit}


The Higgs sectors of the models in the previous subsection have been comprehensively studied. However, here we wish to map them onto the two Higgs doublet model once the adjoint scalars have been integrated out. The standard parametrisation of the Two-Higgs doublet model is
\begin{eqnarray}
V_{EW} &=& m_{11}^2 \Phi_1^\dagger \Phi_1 + m_{22}^2 \Phi_2^\dagger \Phi_2 - [m_{12}^2 \Phi_1^\dagger \Phi_2 + \text{h.c}] + \frac{1}{2} \lambda_1 (\Phi_1^\dagger \Phi_1)^2 + \frac{1}{2} \lambda_2 (\Phi_2^\dagger \Phi_2)^2 \nonumber \\
& & +  \lambda_3(\Phi_1^\dagger \Phi_1) (\Phi_2^\dagger \Phi_2) + \lambda_4 (\Phi_1^\dagger \Phi_2)(\Phi_2^\dagger \Phi_1) \nonumber \\ 
& & + \left[ \frac{1}{2} \lambda_5 (\Phi_1^\dagger \Phi_2)^2 + [\lambda_6 (\Phi_1^\dagger \Phi_1) + \lambda_7 (\Phi_2^\dagger \Phi_2)] \Phi_1^\dagger \Phi_2 + \text{h.c} \right]\,,  \label{reparam2HDM} \,
\end{eqnarray}
To map our supersymmetric model onto this, we choose to make the identification
\begin{align}
\Phi_2 = H_u, \qquad \Phi_1^i = -\epsilon_{ij} (H_d^j)^* \leftrightarrow \twv[H_d^0,H_d^-] = \twv[\Phi_1^0,-(\Phi_1^+)^*] 
\end{align}
from which we can write down
\begin{eqnarray}
m_{11}^2 &=& m_{H_{d}}^2 + \mu^2, \qquad m_{22}^2 = m_{H_{u}}^2 + \mu^2, \qquad m_{12}^2 = B_\mu . 
\label{2HDM_params}
\end{eqnarray}
The parameters $\lambda_i$ were given at tree-level and with some loop corrections in \cite{Belanger:2009wf,Benakli:2012cy} in the limit of neglecting $\mu$ and $m_{DY}, m_{D2}$. However, when we integrate out the adjoint scalars and retain these terms, there are corrections due to the presence of trilinear couplings; setting the parameters $A_S, A_T$ to zero, we find for the minimal model:
\begin{align}
\lambda_1 =& \frac{1}{4} (g_2^2 + g_Y^2) -\frac{\left(g_Y m_{DY} - \sqrt{2} \lambda_S \mu\right)^2}{m_{SR}^2} - \frac{\left(g m_{D2} + \sqrt{2} \lambda_T  \mu\right)^2}{m_{TP}^2}\nn \\
\lambda_2 =&  \frac{1}{4} (g_2^2 + g_Y^2) -\frac{\left(g_Y m_{DY} + \sqrt{2} \lambda_S \mu\right)^2}{m_{SR}^2} - \frac{\left(g m_{D2} - \sqrt{2} \lambda_T  \mu\right)^2}{m_{TP}^2} \nn\\
\lambda_3 =&   \frac{1}{4}(g_2^2 - g_Y^2) + 2 \lambda_T^2 + \frac{g_Y^2 m_{DY}^2 - 2\lambda_S^2 \mu^2}{m_{SR}^2}-  \frac{g^2 m_{D2}^2 - 2\lambda_T^2 \mu^2}{m_{TP}^2} \nn  \\
\lambda_4 =& -\frac{1}{2}g_2^2 + \lambda_S^2 - \lambda_T^2  +\frac{2 g_2^2 m_{D2}^2 - 4 \lambda_T^2 \mu^2}{m_{TP}^2}\,, \nn\\
\lambda_5 =& \lambda_6 = \lambda_7 =0.
\label{EQ:MDGSSMTree}\end{align}
 Here we have defined
\begin{align}
m_{SR}^2 \equiv& m_S^2 + B_S + 4 m_{DY}^2, \qquad m_{TP}^2 \equiv m_T^2 + B_T + 4 m_{D2}^2.
\end{align}
 In fact, the terms suppressed by $m_{SR}, m_{TP}$ all have the effect of suppressing the Higgs quartic coupling: in the limit of large Dirac gaugino masses so that we can neglect $m_S^2, B_S, m_T^2, B_T$ we find
\begin{align}
\lambda_1, \lambda_2 \rightarrow 0, \qquad \lambda_3 \rightarrow 2 \lambda_T^2, \qquad \lambda_4 \rightarrow \lambda_S^2 - \lambda_T^2.
\end{align}
This simply corresponds to the well-known fact (see e.g. \cite{Fox:2002bu}) that the adjoint scalars eliminate the D-term potential of the Higgs, because they couple via the D-term. Writing $\phi_i $ for (anti)fundamental scalars and $\Sigma$ for adjoint scalars, we have
\begin{align}
\lagr \supset& \sqrt{2} m_{D\Sigma} \Sigma^a D^a + g D^a \phi_i^* T^a \phi_i \rightarrow V_D  = \frac{1}{2} \left( \sqrt{2} m_{D\Sigma} \Sigma^a  + g \phi_i^* T^a \phi_i \right)^2 
\end{align}
where $T^a$ are the generators of the gauge group with coupling $g$, and we see that the above
will always be zero when we integrate out $\Sigma$. 

For the MRSSM, for simplicity again neglecting $A_S, A_T$ -- for completeness we give the full corrections in appendix \ref{APP:MRSSM_TreeLevel} -- we find 
\begin{align}
 \lambda_1^{\rm MRSSM} =& \frac{1}{4} (g_2^2 + g_Y^2) - \frac{ (g_Y m_{DY} - \sqrt{2} \lambda_{S_d} \mu_d)^2}{m_{SR}^2} - \frac{ (g_2 m_{D2} + \sqrt{2} \lambda_{T_d} \mu_d)^2}{m_{TP}^2} \nn\\
 \lambda_2^{\rm MRSSM} =& \frac{1}{4} (g_2^2 + g_Y^2) - \frac{ (g_Y m_{DY} + \sqrt{2} \lambda_{S_u} \mu_u)^2}{m_{SR}^2} - \frac{ (g_2 m_{D2} + \sqrt{2} \lambda_{T_u} \mu_u)^2}{m_{TP}^2} \nn\\
 \lambda_3^{\rm MRSSM} =&  \frac{1}{4}(g_2^2 - g_Y^2) \nn\\
&+ \frac{ (g_Y m_{DY} - \sqrt{2} \lambda_{S_d} \mu_d)(g_Y m_{DY} + \sqrt{2} \lambda_{S_u} \mu_u)}{m_{SR}^2}- \frac{ (g_2 m_{D2} + \sqrt{2} \lambda_{T_d} \mu_d)(g_2 m_{D2} + \sqrt{2} \lambda_{T_u} \mu_u)}{m_{TP}^2} \nn\\
 \lambda_4^{\rm MRSSM} =& -\frac{1}{2}g_2^2+ 2\frac{ (g_2 m_{D2} + \sqrt{2} \lambda_{T_d} \mu_d)(g_2 m_{D2} + \sqrt{2} \lambda_{T_u} \mu_u)}{m_{TP}^2} \nn\\
 \lambda_5^{\rm MRSSM}  =&  \lambda_6^{\rm MRSSM}  =  \lambda_7^{\rm MRSSM}  = 0.
\end{align}
In this case, the supersoft limit is even worse, because in that limit \emph{all} of the $\lambda_i$ vanish. However, even with the additions of $\lambda_S$ and $\lambda_T$ in the minimal model, the potential is not stable in this limit -- for example if $H_d$ or $H_u$ are set to zero the quartic terms vanish -- and so we would require loop corrections to prevent runaway vacua. An investigation of whether this is even viable is beyond the scope of this paper: instead, since we do not want to substantially reduce the Higgs quartic coupling at low scales we shall consider instead that $|m_{DY}| \ll m_S, |m_{D2}| \ll m_T$. As is also well known (see e.g. \cite{Benakli:2011kz,Benakli:2012cy}) and we shall later discuss, this limit is also imposed on us by electroweak precision tests. In this limit we have instead at tree-level
\begin{align}
\lambda_1 ,  \lambda_2 \rightarrow  \frac{1}{4} (g_2^2 + g_Y^2), \qquad \lambda_3 \rightarrow  \frac{1}{4} (g_2^2 - g_Y^2)+ 2 \lambda_T^2, \qquad \lambda_4 \rightarrow  -\frac{1}{2} g_Y^2 +\lambda_S^2 - \lambda_T^2,
\label{EQ:DGLambdasTree}\end{align}
and $\lambda_i^{\rm MRSSM} \rightarrow \lambda_i^{MSSM}$:
\begin{align}
\lambda_1^{\rm MSSM} , \lambda_2^{\rm MSSM} \rightarrow  \frac{1}{4} (g_2^2 + g_Y^2), \qquad \lambda_3^{\rm MSSM} \rightarrow  \frac{1}{4} (g_2^2 - g_Y^2), \qquad \lambda_4^{\rm MSSM} \rightarrow  -\frac{1}{2} g_Y^2.
\end{align}

Hence for the rest of the paper we shall consider our low-energy theory to be a type-II two Higgs doublet model with an additional (Dirac) bino and wino (the gluino must remain heavy due to LHC constraints -- currently of the order of $2$ TeV). We shall fix the boundary conditions at high energies and find some interesting conclusions.

\subsection{Tree-level alignment}
\label{Alignment_heavy}

In \cite{Antoniadis:2006uj,Ellis:2016gxa} the Higgs sector of Dirac gaugino models was investigated in the limit that the couplings $\lambda_S, \lambda_T$ took their $N=2$ supersymmetric values at the low energy scale. However, they also pointed out that alignment in the Higgs sector would be broken by quantum corrections to the $(2,2)$ element of the Higgs mass matrix. In this section we shall consider just the potential at tree-level, and in section \ref{SEC:PRECISION} consider loop corrections, contrasting our results with theirs. 

To begin with, the mass-matrices for the CP-even neutral scalars in the two-Higgs doublet model can be parametrised in the \emph{alignment basis} where 
\begin{align}
 \twv[\mathrm{Re}(\Phi_1),\mathrm{Re}(\Phi_2)] = \frac{1}{\sqrt{2}}\twomat[c_\beta,-s_\beta][s_\beta,c_\beta]\twv[v+h,H]
\end{align}
is (see e.g. \cite{Gunion:1989we,Davidson:2005cw, Bernon:2015qea})
\begin{eqnarray}
\mathcal{M}^2_h = \begin{pmatrix}
Z_1 v^2 & Z_6 v^2 \\
Z_6 v^2 & m_A^2 + Z_5 v^2 \end{pmatrix} \,,  
\label{2HDM_mass_matrix}
\end{eqnarray}
where, using $\lambda_{345} \equiv \lambda_3 + \lambda_4 + \lambda_5$ we have 
\begin{align}
Z_1 \equiv& \lambda_1c_\beta^4 + \lambda_2 s_\beta^4 + \frac{1}{2} \lambda_{345} s_{2\beta}^2, \qquad Z_5 \equiv \frac{1}{4} s_{2\beta}^2 \left[ \lambda_1 + \lambda_2 - 2\lambda_{345}\right] + \lambda_5 \nn\\
Z_6 \equiv& -\frac{1}{2} s_{2\beta} \left[\lambda_1 c_\beta^2 - \lambda_2 s_\beta^2 - \lambda_{345} c_{2\beta} \right].
\label{EQ:THDMZis}\end{align} 
The parameter $m_A$ is the pseudoscalar mass, given by
\begin{align}
m_A^2 =& - \frac{m_{12}^2}{s_\beta c_\beta} - \lambda_5 v^2,
\end{align}
while the charged Higgs mass is
\begin{align}
m_{H^+}^2 =& \frac{1}{2} ( \lambda_5 - \lambda_4) v^2 + m_{A}^2. 
\end{align}
The neutral Higgs masses are 
\begin{eqnarray} 
m_{H,h}^2 &=& \frac{1}{2} \left[m_A^2 + (Z_1 + Z_5)v^2 \pm \sqrt{ \left(m_A^2 + (Z_5 - Z_1)v^2\right)^2 + 4Z_6^2 v^4} \, \right].
\end{eqnarray} 
For our \emph{minimal} model we have
\begin{eqnarray}
Z_1 &=& \frac{1}{4} (g_2^2 + g_Y^2) (1-s_{2\beta}^2) + \frac{s_{2\beta}^2}{2} (\lambda_S^2 + \lambda_T^2) \label{Z_1_Specific}\\
Z_5 &=& \frac{1}{2} s_{2\beta}^2 \left[ \frac{(g_2^2 + g_Y^2)}{2} - (\lambda_S^2 + \lambda_T^2)\right]\\ 
Z_6 &=& -\frac{1}{2} s_{2\beta} c_{2\beta} \left[ \frac{(g_2^2 + g_Y^2)}{2} - (\lambda_S^2 + \lambda_T^2)\right]\, \label{Z_6_Specific}
\end{eqnarray} and
\begin{eqnarray}
m_{H,h}^2 &=& \frac{1}{2} \left[ m_A^2 + \frac{v^2}{4} (g_2^2 + g_Y^2) \pm v^2 \left[ \left(\frac{1}{4} (g_2^2 + g_Y^2)(2 s_{2\beta}^2 -1) - s_{2\beta}^2 (\lambda_S^2 + \lambda_T^2) + \frac{m_A^2}{v^2} \right)^2 \right. \right. \nonumber \\
& & \left. \left. + s_{2\beta}^2 c_{2\beta}^2 \left(  \frac{(g_2^2 + g_Y^2)}{2} - (\lambda_S^2 + \lambda_T^2)\right)^2 \right]^{1/2} \right]\,.
\end{eqnarray}
 The Higgs mass matrix is diagonalised to find the physical Higgs masses and the mixing angle $\alpha$. 
From the identification of the 2HDM parameters in \eqref{2HDM_params} we obtain
\begin{align}
s_{2(\beta - \alpha)} =& \frac{v^2}{m_H^2 - m_h^2} s_{2\beta} c_{2\beta} \bigg[ \frac{(g_2^2 + g_Y^2)}{2} - (\lambda_S^2 + \lambda_T^2)\bigg], \nn\\
c_{\beta - \alpha} =& \frac{s_{2\beta} c_{2\beta} \, v^2 \left( g_Y^2 + g_2^2 - 2(\lambda_S^2 + \lambda_T^2)\right)}{4 \sqrt{\left(m_H^2 - m_h^2\right)\left(m_H^2 - \frac{v^2}{2}\Bigl\{(g_Y^2 + g_2^2)\frac{c^2_{2\beta}}{2} + (\lambda_S^2 + \lambda_T^2)s^2_{2\beta}\Bigr\}\right)}} \,. \label{c_ba}
\end{align} 
The condition for alignment is the diagonalisation of $\mathcal{M}^2$ i.e. $Z_6 \rightarrow$ 0. From equation \eqref{Z_6_Specific} we see this amounts at tree-level to having
\begin{eqnarray}
\lambda_S^2 + \lambda_T^2 = \frac{g_Y^2 + g_2^2}{2} \label{N=2}.
\label{Alignment_couplings}
\end{eqnarray} 
In other words, when the couplings respect their $N=2$ values, the Higgs doublets are \emph{automatically aligned}! 
From equations (\ref{c_ba}, \ref{N=2}) we find that in this alignment limit, $c_{\beta - \alpha} \rightarrow 0$ and  $s_{\beta - \alpha} \rightarrow 1$, therefore  the heavy CP-even neutral scalar doest not take part in electroweak symmetry breaking while $h$ is a Standard Model Higgs-like boson. The tree-level masses of the two neutral CP-even Higgs bosons  are
\begin{eqnarray}
m_h^{N=2} = m_Z \,, \;\;\;\;\;\; m_H^{N=2} = m_A\,,  \label{hH_Aligned}
\end{eqnarray}
while the charged Higgs boson mass is given by
\begin{eqnarray}
m^{2, N=2}_{H^{\pm}} = m_A^2 + 3m_W^2 - m_Z^2 , \label{H_charged_Aligned}
\end{eqnarray}
correcting the expression given in \cite{Antoniadis:2006uj,Ellis:2016gxa}.
Hence, at tree-level, the model exhibits alignment \emph{for any value of $\tan \beta$} and the tree-level Higgs mass is independent of $\tan \beta$ (which was already noted in \cite{Antoniadis:2006uj,Ellis:2016gxa}).

On the other hand, for the MRSSM there is no automatic alignment, because the Higgs sector at tree-level closely resembles that of the MSSM once the adjoint scalars and R-Higgs fields are decoupled; this can be seen just by putting $\lambda_S = \lambda_T =0$ in the above equations. In the following we shall therefore mostly focus on the minimal Dirac gaugino model (with some further comments about the MRSSM).

\section{Radiative corrections to alignment}
\label{SEC:RADIATIVE}

As mentioned above, the perfect alignment obtained at tree-level is not preserved when the radiative corrections to the scalar effective potential are taken into account. In addition to the corrections already present in the MSSM, there are two new sources for this misalignment. The first is due to the appearance of chiral fields, quarks and leptons, at a scale $M_{N=2}$. 
This scale can be identified with the fundamental scale of the theory, or an intermediate scale where a partial breaking  $N=2 \rightarrow N=1$ is achieved (while an explicit realisation  of this  partial  supersymmetry breaking remains unknown for a chiral theory, there is not a no-go theorem showing it to be impossible). 
The second large contribution comes from the mass splitting between fermonic and bosonic components of all of the superfields, i.e. coming from the $N=2 \rightarrow N=0$ (or $N=1 \rightarrow N=0$) breaking. We will discuss them here in turn. 

\subsection{Misalignment from $N=2 \rightarrow N=1$ (chiral matter)}

When we run our couplings from the $N=2$ scale $M_{N=2}$ to the scale of the $N=1$ supersymmetric superparticles (which we shall call $\MSUSY$) there will be a splitting induced of $\lambda_S$ and $\lambda_T$ relative to the $N=2$ SUSY relations. This in turn will lead to misalignment at $\MSUSY$ via a non-zero $Z_6$:
\begin{align}
Z_6 (\MSUSY) =& \frac{1}{4} s_{2\beta} c_{2\beta} \bigg[ (2\lambda_S^2 - g_Y^2) +  (2\lambda_T^2 - g_2^2) \bigg] + \mathrm{threshold\ corrections}.
\end{align} 
To obtain an estimate of the magnitude of this splitting, we can integrate over the difference in the beta functions for $\lambda_S$ and $\lambda_T$ to leading order:
\begin{eqnarray}
\left[2 \lambda_S^2 - g_Y^2\right]_{\MSUSY} &=& -\frac{2 g_Y^2}{16 \pi^2} \left[3|y_t|^2 + 3|y_b|^2 + |y_\tau|^2 - 10 g_Y^2 \right] \log\left(\frac{M_{N=2}}{\MSUSY} \right)\,, \label{2ls} \\
\left[2 \lambda_T^2 - g_2^2\right]_{\MSUSY} &=& -\frac{2 g_2^2}{16 \pi^2} \left[ 3|y_t|^2 + 3|y_b|^2 + |y_\tau|^2 - 6 g_2^2 \right] \log\left(\frac{M_{N=2}}{\MSUSY} \right) \label{2lt} \,.
\end{eqnarray}
These equations are only useful for small $M_{N=2}/\MSUSY$, because for large ratios the top Yukawa coupling can change by a factor of two or more, but it gives an indication of the amount of misalignment: even for $\log\left(\frac{M_{N=2}}{\MSUSY} \right) \sim \mathcal{O}(10) $ we find
 \begin{align}
Z_6 (\MSUSY) \sim & - \mathcal{O}(0.1) \frac{t_\beta}{1+t_\beta^2} \left(\frac{t_\beta^2 - 1}{t_\beta^2 +1} \right) \left(\frac{\log M_{N=2}/\MSUSY}{10} \right).
\end{align}
This is a small deviation from alignment indeed, and very encouraging.
We shall investigate this quantitatively in section \ref{SEC:PRECISION}, and will find that due to the diminishing Yukawa couplings at high energies\footnote{This is true for reasonable values of $\tan \beta \gtrsim 1.5$. For values of $\tan \beta$ near unity the Yukawa couplings diverge at high energies so we cannot consistently place our $N=2$ scale there.}, the actual splitting is smaller than this naive estimate. As an aside, a similar conclusion is reached if we extend our Dirac Gaugino theory by including additional fields to the Minimal Dirac Gaugino Supersymmetric standard Model (MDGSSM)  to restore gauge coupling unification.


\subsection{Misalignment  from $N=1 \rightarrow N=0$ (mass splitting)}
\label{MassSplit}
 

More significantly, there is the potential misalignment induced from the threshold corrections at $\MSUSY$ and then the running between $\MSUSY$ and the scale of the THDM; let us take the matching scale (as commonly done) to be the electroweak vev $v$. These can both be approximated at one loop by corrections to the $\delta \lambda_i$. In the approximation that the singlet and triplet scalars are degenerate with mass $M_\Sigma$, and the stop squarks are degenerate with mass $m_{\tilde{t}}$ and neglecting the splitting between the couplings $\lambda_{S,T}$ and their $N=2$ values we find, matching at a scale $\mu$:
\begin{align}
\delta \lambda_1 =&\frac{1}{16\pi^2}  \log \frac{M_{\Sigma}^2 }{\mu^2}\bigg[ \lambda_S^4 + 3 \lambda_T^4   + 2 \lambda_{S}^2 \lambda_T^2  \bigg] \nn\\
 \delta  \lambda_2 =& \delta \lambda_1 + \frac{3y_t^4}{8\pi^2} \log \frac{m_{\tilde{t}}^2}{\mu^2} \nn\\
\delta \lambda_3 =& \frac{1}{16\pi^2} \log \frac{M_{\Sigma}^2 }{\mu^2}\bigg[ \lambda_S^4  + 3\lambda_T^4   - 2 \lambda_{S}^2 \lambda_T^2  \bigg] \nn\\
\delta \lambda_4 =&\frac{1}{16\pi^2}   4 \lambda_S^2 \lambda_T^2 \log \frac{M_{\Sigma}^2 }{\mu^2},
\label{EQ:simpleST}\end{align}
using $y_t, y_b, y_\tau$ to denote the top, bottom and $\tau$ Yukawa couplings.
We give full (updated) expressions in the limit $m_{DY},m_{D2} \ll m_S, m_T$ in appendix \ref{APP:THRESHOLDS}.

We then find the remarkable result that the singlet/triplet scalar contributions to $Z_6$ \emph{exactly cancel out}! We then find that the dominant contribution to $Z_6$ is that coming from the stops: 
\begin{align}
Z_6(v) \simeq& Z_6 (\MSUSY) + s_\beta^3 c_\beta \times  \frac{3y_t^4}{8\pi^2} \log \frac{m_{\tilde{t}}^2}{m_t^2} ,
\end{align}
where $m_t$ is the top quark mass. 
Although the magnitude of this is the same as the loop contribution to $Z_6$ in the MSSM, the misalignment thus induced is much smaller, because (a) there is no tree-level contribution, and (b) it is also proportional to the stop correction to the Higgs mass, which is smaller than in the MSSM due to the tree-level boost to the Higgs mass. To investigate the misalignment in this model further, however, we shall in the next section perform a precision study using numerical tools, where we shall use the logic of the hMSSM\cite{Djouadi:2013uqa}/h2MSSM \cite{Ellis:2016gxa} to show that the misalignment in the model is even smaller than the above naive estimate.

\section{Precision study}
\label{SEC:PRECISION}

To precisely study the quantum corrections to alignment in our minimal model, we implemented the low-energy model consisting of the THDM supplemented by a Dirac bino and a Dirac Wino into the package {\tt SARAH}. We describe the couplings of the model in detail in appendix \ref{APP:THDMEW}. We then modified the code to implement the boundary at a supersymmetry scale $M_{\rm SUSY}$ and use the two-loop supersymmetric RGEs for the minimal Dirac gaugino extension of the MSSM \cite{Goodsell:2012fm,Arvanitaki:2013yja} as generated by \SARAH \cite{Staub:2008uz,Staub:2012pb,Staub:2013tta}. While this theory does not fit into a GUT and has no gauge coupling unification, we implemented an $N=2$ supersymmetry scale where the couplings $\lambda_S, \lambda_T$ take their $N=2$ supersymmetric values. By running all the way from a low-scale $Q$ (which we take to be the scale of the Dirac gauginos and Heavy Higgses, but could equally be $m_{\rm top}$) up to $M_{\rm SUSY}$ and then $M_{N=2}$ and back down, iterating until the results converge, we were able to find consistent values of the parameters. At the scale $Q$, the threshold corrections are those that are included in {\tt SARAH} by default:
\begin{itemize}
\item One-loop matching of Yukawa couplings to the Standard Model values, plus two-loop strong corrections to the top Yukawa.
\item One-loop gauge threshold corrections.
\item Two-loop corrections to the Higgs masses \cite{Goodsell:2014bna,Goodsell:2015ira,Braathen:2017izn} (which implement the generic expressions of \cite{Martin:2001vx,Martin:2003it,Goodsell:2015ira} and the solution to the Goldstone Boson catastrophe of \cite{Braathen:2016cqe,Braathen:2017izn}).
\end{itemize}
We employed the two-loop RGEs for this model up to \MSUSY, and then at the scale $M_{\rm SUSY}$, we implemented the following thresholds:
\begin{itemize}
\item Tree-level correction to the $\lambda_i$ from Dirac gaugino masses given in (\ref{EQ:MDGSSMTree}), even if we are otherwise neglecting the Dirac gaugino masses.
\item One-loop corrections to the $\lambda_i$ given in \ref{app:ST}.
\item Conversion of \MS to \DR gauge couplings given in \ref{app:MSDR}.
\item Conversion of \MS to \DR Yukawa couplings proportional to the strong gauge coupling, given in \ref{app:MSDR}.
\end{itemize}
We take \MSUSY to be a common mass of left- and right-handed stops, and assume that other MSSM particles have masses at this scale; we allow the singlet at triplet scalars to be heavier at a scale $M_{\Sigma}$. We eliminate all R-symmetry-violating terms (such as squark trilinear couplings) and assume that
$$m_{DY}, m_{D2}, \mu \ll \MSUSY.$$
This means that we neglect squark mixing, which greatly simplifies the thresholds. 
The thresholds for supersymmetric particles that we include are then \emph{nearly} complete in this limit: the gauge and Yukawa threshold corrections vanish for the MSSM couplings, and we only neglect the corrections to the gauge/Yukawas induced by the adjoint scalars -- since their effect is in general very small; we leave the calculation and implementation of these for future work. However, we do include their contribution to the Higgs quartic couplings. Furthermore, we know that in the limit of zero squark mixing the two-loop corrections to the Higgs quartic couplings are also small or even vanishing \cite{Bagnaschi:2014rsa}, and so we are justified in neglecting them. 

To perform a more general scan over the parameter space including trilinear scalar couplings, general masses and allowing $\mu, m_{DY}, m_{D2}$ to be of the order of \MSUSY  we would need to compute the additional threshold corrections. While we expect that the effect of $\mu, m_{DY}, m_{D2} $ on our results will be very small, it would nonetheless be interesting to compute these in the future.

We performed scans over the values of $\tan \beta$ and varied \MSUSY to obtain a light Higgs mass of $125.15$ GeV. For the other values we take 
\begin{align}
M_{\Sigma} = 5\ \TeV, \qquad (m_{DY}, m_{D2}, \mu) = (400, 600, 500)\ \GeV, \qquad m_A^{\rm tree} = 600\ \GeV
\label{EQ:scanvalues}\end{align}
 by imposing
\begin{align}
m_{12}^2 =& - (m_A^{\rm tree})^2 s_\beta c_\beta. 
\end{align}
As we shall later see, these are compatible with all current experimental constraints. Note, on the other hand, that we shall not discuss collider limits on the electroweakinos because the effect of changing their masses is tiny.  

In the scans we see little deviation between $m_A^{\rm tree} $ and the mass of the heavy/charged Higgses because the mixing is small; indeed the results are not especially sensitive to $m_A^{\rm tree}$ as a result.

\subsection{Running from the $N=2$ scale}

\begin{figure}
\centering
\includegraphics[width=0.8\textwidth]{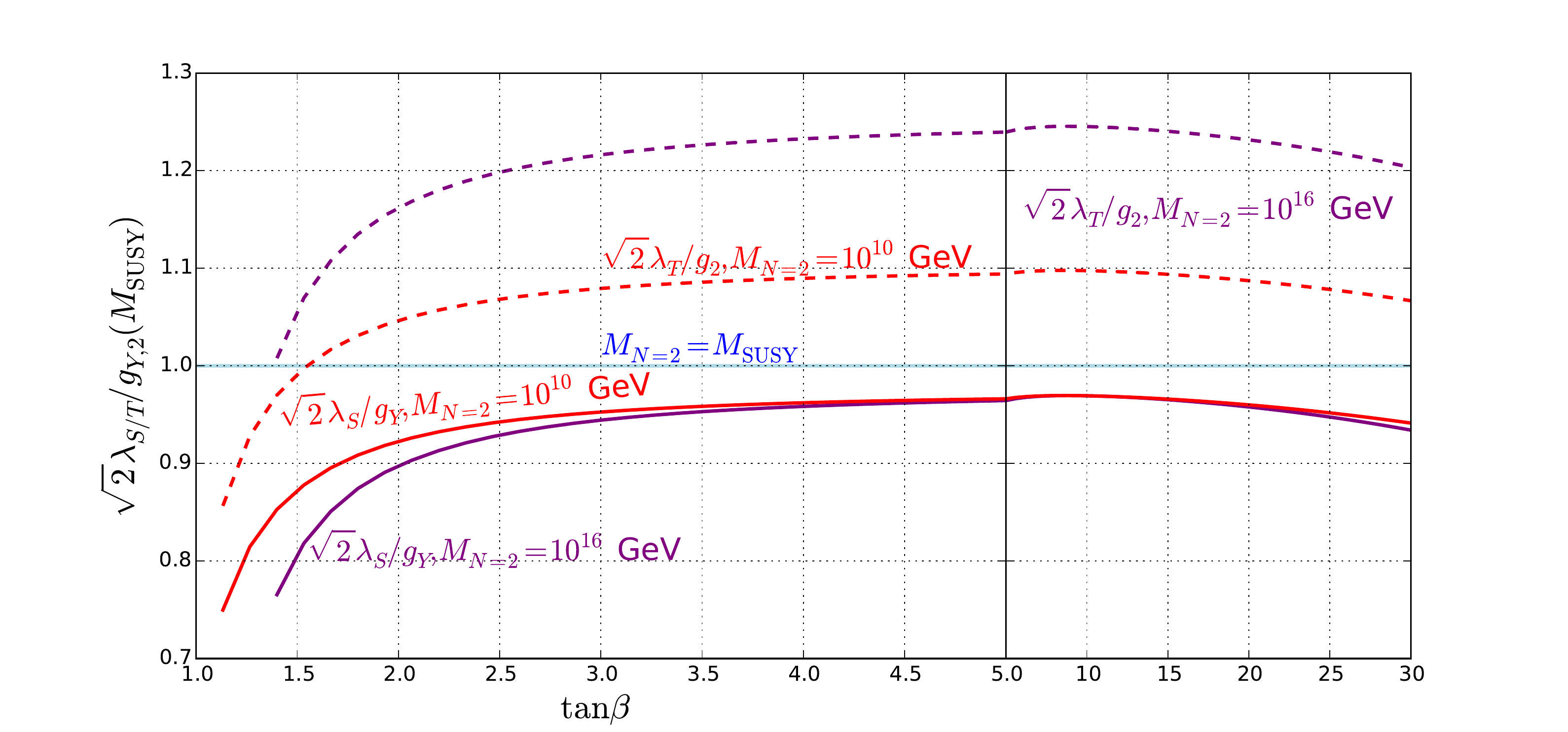}
\caption{Variation of the ratios $\sqrt{2}\lambda_S/g_Y$ and $\sqrt{2}\lambda_T/g_2$ at the scale \MSUSY with $\tan \beta$, for $M_{N=2} = \MSUSY, 10^{10}\ \GeV$ and $10^{16}$ GeV.}
\label{FIG:MSUSYcouplings}\end{figure}

At the scale $Q=400$ GeV, we find $g_Y = 0.37, g_2= 0.64 \pm 0.01$. These are barely different at the SUSY scale and vary little with $M_{N=2}$, but we do find some dependence of the ratios $\sqrt{2} \lambda_S/g_Y, \sqrt{2} \lambda_T/g_2$ on this scale, which we give in figure \ref{FIG:MSUSYcouplings}. The values in the plot were taken with a common supersymmetric scale of $M_{\rm SUSY} =3$ TeV and have essentially no dependence on $m_A$.

\begin{figure}
\centering
\includegraphics[width=0.6\textwidth]{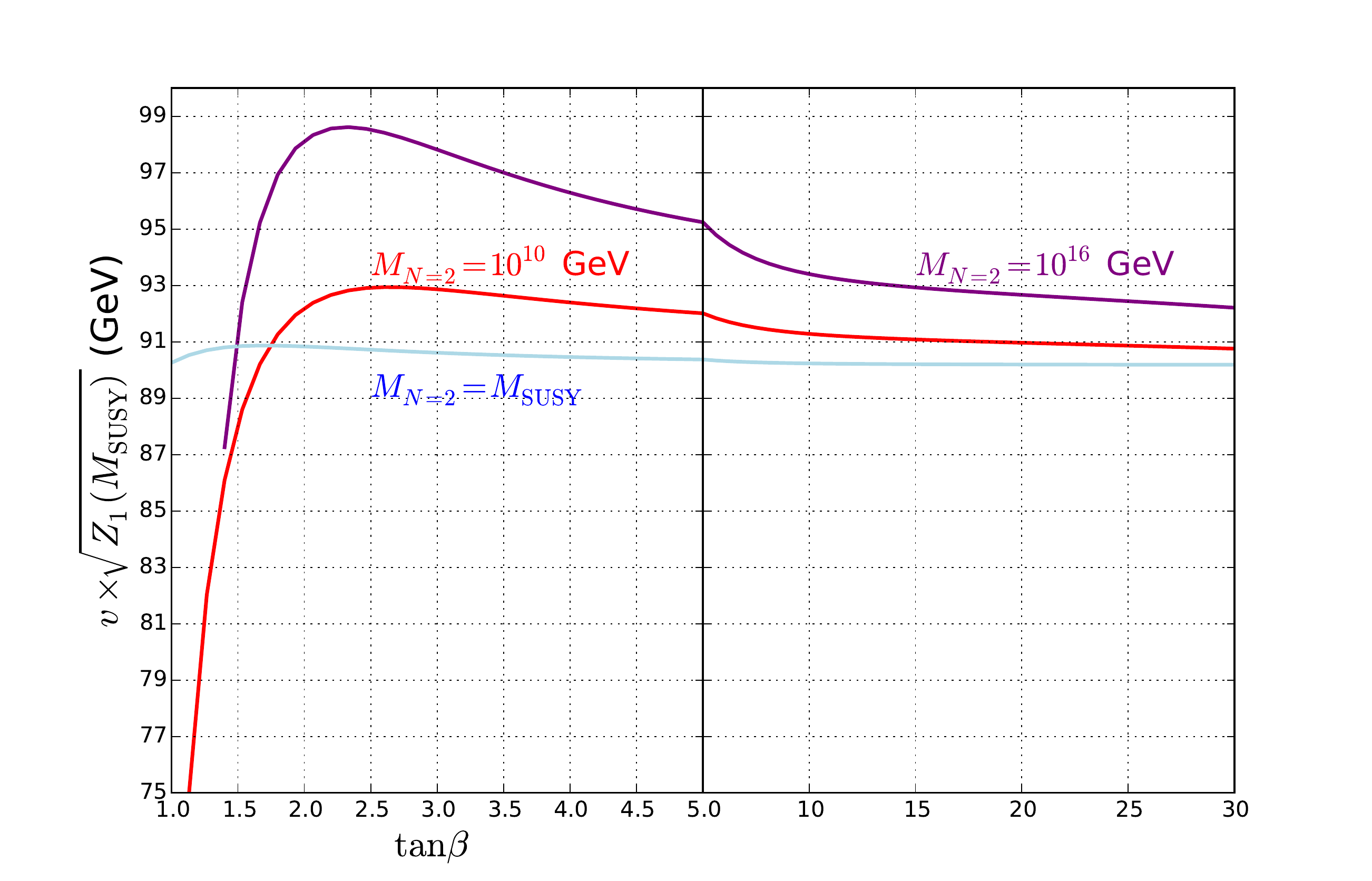}
\caption{$v\times \sqrt{Z_1(\MSUSY)}$ against $\tan \beta$ for $M_{N=2} = \MSUSY, 10^{10}$ GeV and $10^{16}$ GeV, which corresponds to the ``tree-level'' value of the Higgs mass before we take running from $\MSUSY$ (or equivalently the SUSY corrections at $M_Z$) into account (we take $v=246$ GeV in the figure). We see that increasing $M_{N=2}$ \emph{increases} the Higgs mass, particularly for small $\tan \beta > 1.5$.}
\label{FIG:MSUSYZ1}\end{figure}

An alternative way of visualising this information is in the quantity $Z_1$ evaluated at the SUSY scale. Since our model is always very near alignment, this gives the ``tree-level'' Higgs mass and so in figure \ref{FIG:MSUSYZ1} we plot $v \sqrt{Z_1 (M_{\rm SUSY})}$. We see that for $M_{N=2} = M_{\rm SUSY}$ this is always essentially $M_Z$, while as we increase $M_{N=2}$ we obtain a \emph{further} enhancement to the Higgs mass at small $\tan \beta \gtrsim 1.5$.

\begin{figure}\centering
\includegraphics[width=0.7\textwidth]{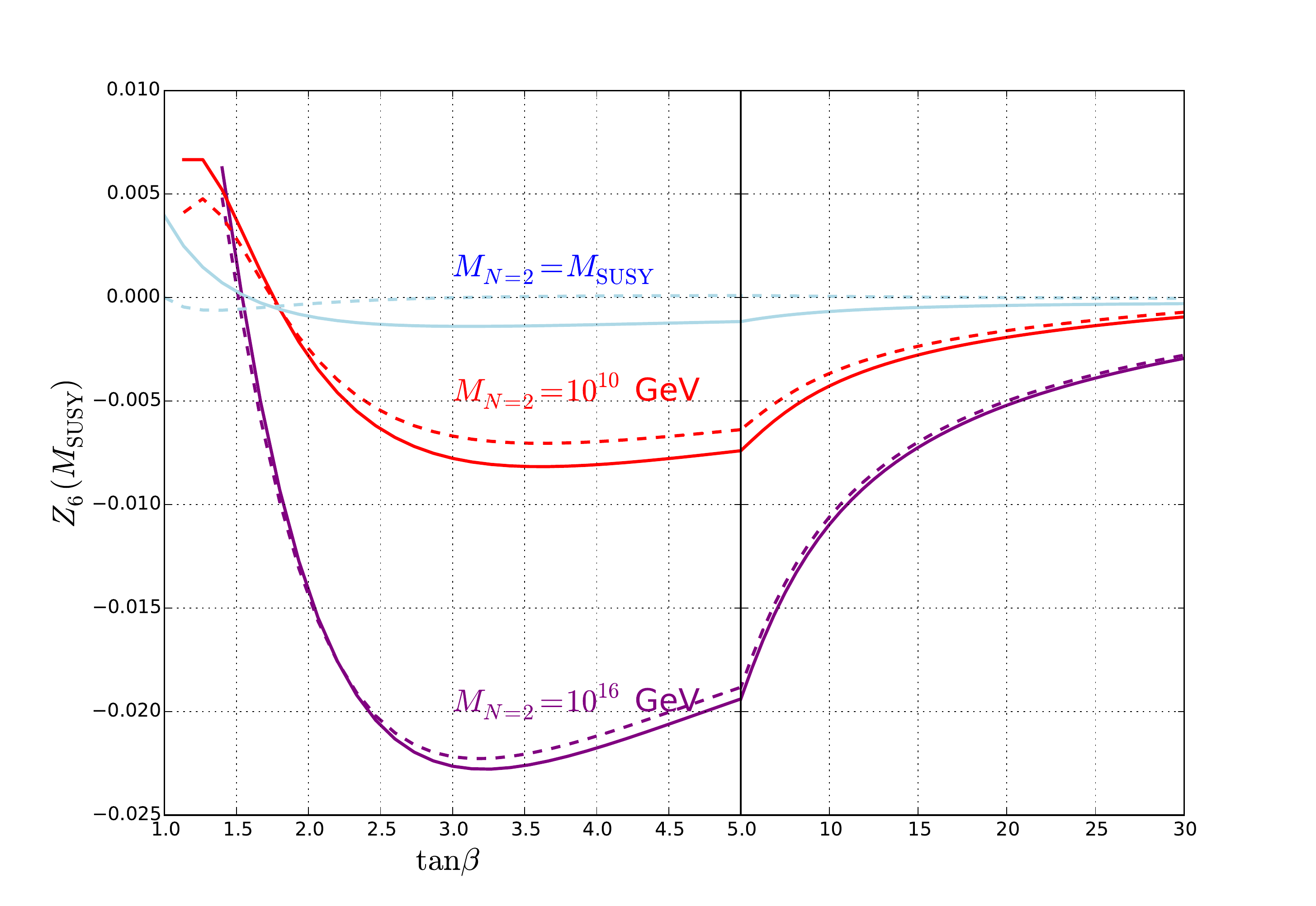}
\caption{$Z_6 (\MSUSY)$ against $\tan \beta$ for $M_{N=2} = \MSUSY, 10^{10}\ \GeV$ and $10^{16}$ GeV, which corresponds to just the contributions to $Z_6$ from the running of $\lambda_{S,T}$ and the threshold corrections. The solid lines show the full value of $Z_6$, while the dashed lines are just those given by equation (\ref{EQ:Z6SUSY}), i.e. without threshold corrections. }
\label{FIG:Z6MSUSY}\end{figure}

If we were to include no further corrections, then the value of $Z_6$ at \MSUSY would be given by
\begin{align}
Z_6 (\MSUSY) =& \frac{1}{4} s_{2\beta} c_{2\beta} \bigg[ g_Y^2 (2\lambda_S^2/g_Y^2 - 1) + g_2^2 (2\lambda_T^2/g_2^2 - 1) \bigg] .
\label{EQ:Z6SUSY}\end{align} 
Crucially then we see that for $M_{N=2} > \MSUSY$ this is dominated by the relative positive shift in $\lambda_T$, which in turn yields a negative contribution to $Z_6$. 
The results from our scans for the value of $Z_{6}$ at the SUSY scale almost exactly correspond to the above equation, which we plot in figure \ref{FIG:Z6MSUSY}. The differences (particularly the tiny difference from zero for the $N=2$ scale equal to \MSUSY) come from the tree-level and loop-level shifts.

\subsection{Running below \MSUSY}

\begin{figure}\centering
\includegraphics[width=0.7\textwidth]{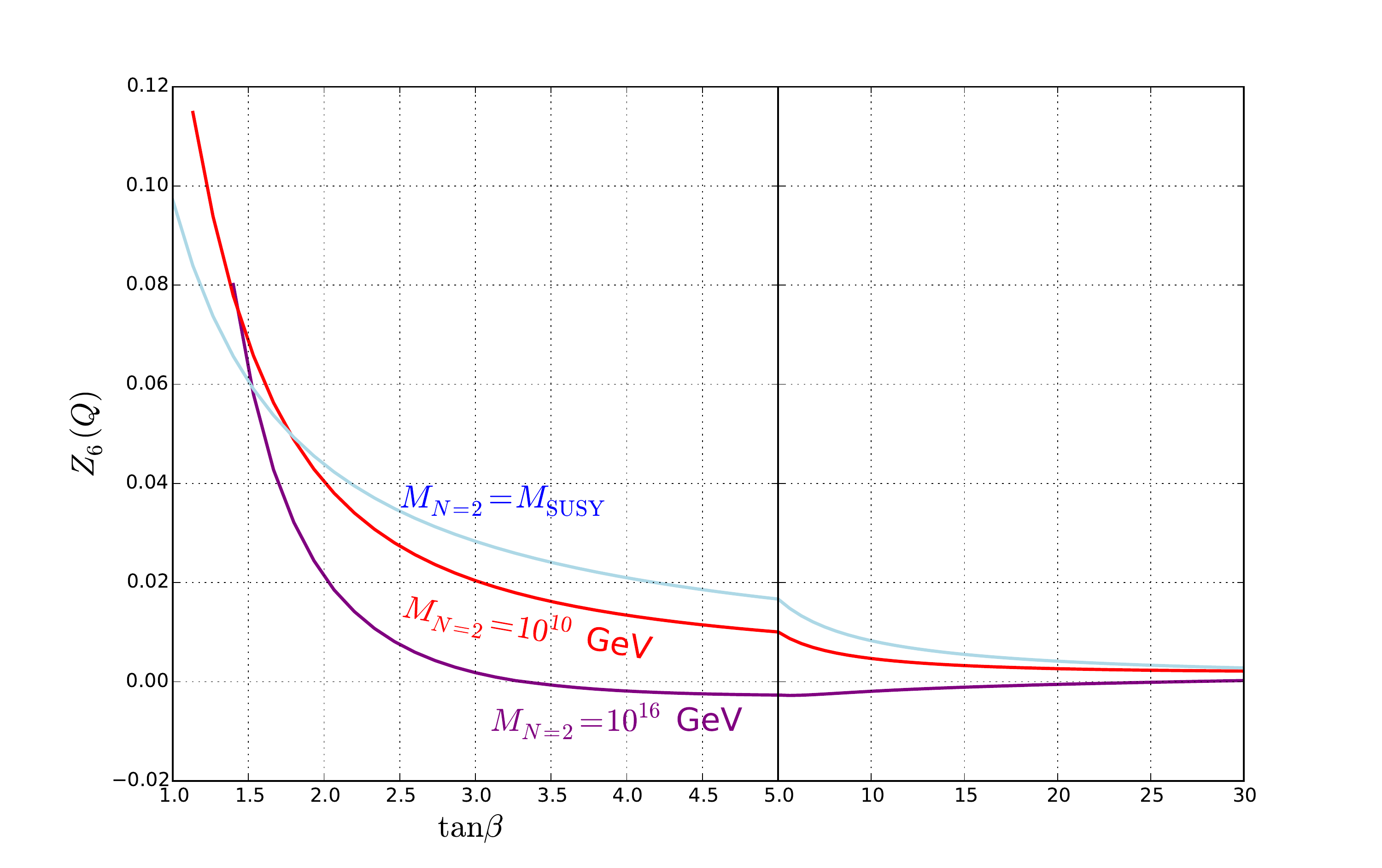}
\caption{$Z_6(Q)$ against $\tan \beta$, where $Q=400$ GeV is our low-energy matching scale. We find that the model shows good alignment for all values of $\tan \beta > 1.5$, with the surprising conclusion that raising the $N=2$ scale \emph{improves} the alignment.}
\label{FIG:Z6}\end{figure}

Once we include the two-loop running below \MSUSY, the picture changes substantially. This is dominated by the effects of the stops via their absence from the RGEs; we plot the results of $Z_6$ for the same scan as in figure \ref{FIG:Z6MSUSY} at the scale of our low-energy theory in figure \ref{FIG:Z6}.

Interestingly, the results can be understood by following the reasoning of the hMSSM\cite{Djouadi:2013uqa}/h2MSSM \cite{Ellis:2016gxa} treatment. In that framework, the quantum corrections to the Higgs mass are assumed to be dominated by the $(2,2)$ component -- and further that we can neglect the contributions to the other components compared to the tree-level ones. We shall first review what happens in the hMSSM and then apply the analysis to our case.

\subsubsection{(Lack of) alignment in the hMSSM}

In the hMSSM \cite{Djouadi:2013uqa}, we have $\lambda_2 = \frac{M_Z^2 + \epsilon}{v^2}  $, where $\epsilon$ encodes the loop corrections (dominated by stops), and all other terms are taken to have their tree-level values, giving the neutral Higgs mass matrix in the alignment basis of
\begin{align}
m_{h,H}^2 =\twomat[M_Z^2 c_{2\beta}^2 + \epsilon s_\beta^4, -M_Z^2 s_{2\beta} c_{2\beta}  + s_\beta^3 c_\beta \epsilon ][-M_Z^2 s_{2\beta} c_{2\beta}  + s_\beta^3 c_\beta \epsilon, m_A^2 + M_Z^2 s_{2\beta}^2 + s_\beta^2 c_\beta^2 \epsilon ].
\end{align}
Now let us suppose that we tune the values to obtain alignment. We then have
\begin{align}
-M_Z^2 s_{2\beta} c_{2\beta}  + s_\beta^3 c_\beta \epsilon =& 0 \nn\\
M_Z^2 c_{2\beta}^2 + \epsilon s_\beta^4 =& m_h^2
\end{align}
which leads to 
\begin{align}
c_{2\beta} =& \frac{m_h^2}{M_Z^2} > 1 \qquad \mathrm{or}\qquad  c_\beta =0,
\end{align}
i.e. it is impossible to achieve alignment without decoupling or going to the large $\tan \beta$ limit with these approximations. If we do not neglect the other contributions to $Z_6$, in the case of exact alignment we then have
\begin{align}
0 = Z_6  =& \frac{1}{v^2 t_\beta} ( m_h^2 - M_Z^2 c_{2\beta}) - s_\beta c_\beta ( c_\beta^2 \delta \lambda_1 - c_{2\beta} \delta \lambda_{345} ) \nn\\
0=&  ( m_h^2 - M_Z^2 c_{2\beta}) - v^2 s_\beta^2 ( c_\beta^2 \delta \lambda_1 - c_{2\beta} \delta \lambda_{345} ) 
\end{align}
Since we expect $\lambda_1, \lambda_{345} \ll \lambda_2$, and for $t_\beta > 1$ we have $s_\beta^2 > c_\beta^2, s_\beta^2 > |c_{2\beta}|$, this is still impossible to satisfy. \footnote{In the full MSSM the radiative corrections to $\lambda_{345}$ and $\lambda_7$ -- which we are neglecting -- can be large enough to allow  alignment for $\tan \beta \gtrsim 10$ \cite{Carena:2013ooa,Carena:2014nza}. However, this also requires significant stop mixing, which we do not have in our model.}. However, we will find that for our scenario things are somewhat better.

\subsubsection{Alignment in the Dirac-gaugino model}

Using the expressions (\ref{EQ:DGLambdasTree}) for the quartic couplings, we can rewrite
\begin{align}
\lambda_1 \equiv& \frac{M_Z^2}{v^2} + \delta \lambda_1 , \qquad \lambda_2 \equiv\frac{M_Z^2}{v^2} + \frac{\epsilon}{v^2}, \nn\\
\lambda_{345} \equiv&  \frac{M_Z^2}{v^2}   + \frac{1}{2} ( 2 \lambda_S^2 - g_Y^2)+ \frac{1}{2} ( 2 \lambda_T^2 - g_2^2) + \delta \lambda_{345}.
\end{align}
This leads to 
\begin{align}
Z_1 v^2 =& M_Z^2 + \epsilon s_\beta^4 + \delta \lambda_1 c_\beta^4 + \frac{1}{2} \delta \lambda_{345} s_{2\beta}^2 + v^2 \bigg[ ( 2 \lambda_S^2 - g_Y^2)+ ( 2 \lambda_T^2 - g_2^2) \bigg] s_\beta^2 c_\beta^2 \nn\\
Z_6 v^2 =& s_\beta^3 c_\beta \epsilon - v^2 s_\beta c_\beta ( c_\beta^2 \delta \lambda_1 - c_{2\beta} \delta \lambda_{345} ) + \frac{1}{2} c_{2\beta} s_\beta c_\beta v^2 \bigg[ ( 2 \lambda_S^2 - g_Y^2)+ ( 2 \lambda_T^2 - g_2^2) \bigg] .
\end{align}
The corrections $\delta \lambda_i$ can be interpreted as either coming from running the couplings between the scale $\MSUSY$ and $Q$, or alternatively from integrating out the supersymmetric particles at the scale $Q$. In the latter case we can obtain an estimate of their values from the expressions (\ref{EQ:simpleST}) and see that they are typically suppressed relative to $\epsilon/v^2$ by a numerical factor and also the ratio of the electroweak gauge coupling to the strong gauge coupling or top Yukawa, and we find that we can therefore continue with the hMSSM approximation and neglect them. However, the effect from the running of $\lambda_S, \lambda_T$ is non-negligible: eliminating $\epsilon$ in exchange for the Higgs mass and defining
\begin{align}
\hat{\delta} \lambda_{345} \equiv \frac{1}{2} ( 2 \lambda_S^2 - g_Y^2)+ \frac{1}{2} ( 2 \lambda_T^2 - g_2^2)
\end{align}
we have
\begin{align}
Z_6  =& \frac{s_\beta c_\beta}{v^2 (m_A^2 s_\beta^2 + M_Z^2 c_\beta^2 - m_h^2)} \bigg[(m_A^2 -m_h^2)(m_h^2 - M_Z^2)  - v^2 \hat{\delta} \lambda_{345} \bigg( m_A^2 s_\beta^2 - M_Z^2 c_\beta^2 + m_h^2 c_{2\beta} \bigg) + v^4 c_\beta^2 s_\beta^2 (\hat{\delta} \lambda_{345})^2\bigg] \nn\\ 
\approx&\frac{0.12}{t_\beta} - \frac{1}{2} \frac{t_\beta}{1+t_\beta^2} \bigg[  ( 2 \lambda_S^2 - g_Y^2)+ ( 2 \lambda_T^2 - g_2^2) \bigg].
\end{align}
We shall later give the expressions for eliminating $\lambda_2$ and calculating $Z_6$ in any THDM with general $\lambda_i, i=1...4$ in equations (\ref{EQ:hTHDM}) and (\ref{EQ:hTHDMZ6}).

A comparison of the above formula with the curves in figure \ref{FIG:Z6} shows that this gives a reasonable fit. In the case of $M_{N=2} = \MSUSY$ the expression is particularly simple, but in the other cases we need to take account of the varation of $\sqrt{2}\lambda_S(\MSUSY), \sqrt{2}\lambda_T(\MSUSY)$ with $\tan \beta$ that can be seen in figure \ref{FIG:MSUSYcouplings}. 

The main conclusion that can be drawn from the above formula is that the misalignment coming from the squark corrections required to enhance the Higgs mass can be compensated by the effect of running $\lambda_S, \lambda_T$. Indeed, we see from figure \ref{FIG:Z6} that for $M_{N=2} = 10^{16}$ GeV, $Z_6$  is essentially vanishing for $\tan \beta \gtrsim 3$. From the curves in the figure, we see that increasing the $N=2$ scale causes a partial or total cancellation of the misalignment contributions, meaning that the Higgs boson is accidentally very Standard-Model-like, independent of the mass of the heavy Higgs! This is the main result of the paper.  

\subsection{Higgs mass bounds on the SUSY scale}

\begin{figure}[ht]\centering
\includegraphics[width=0.6\textwidth]{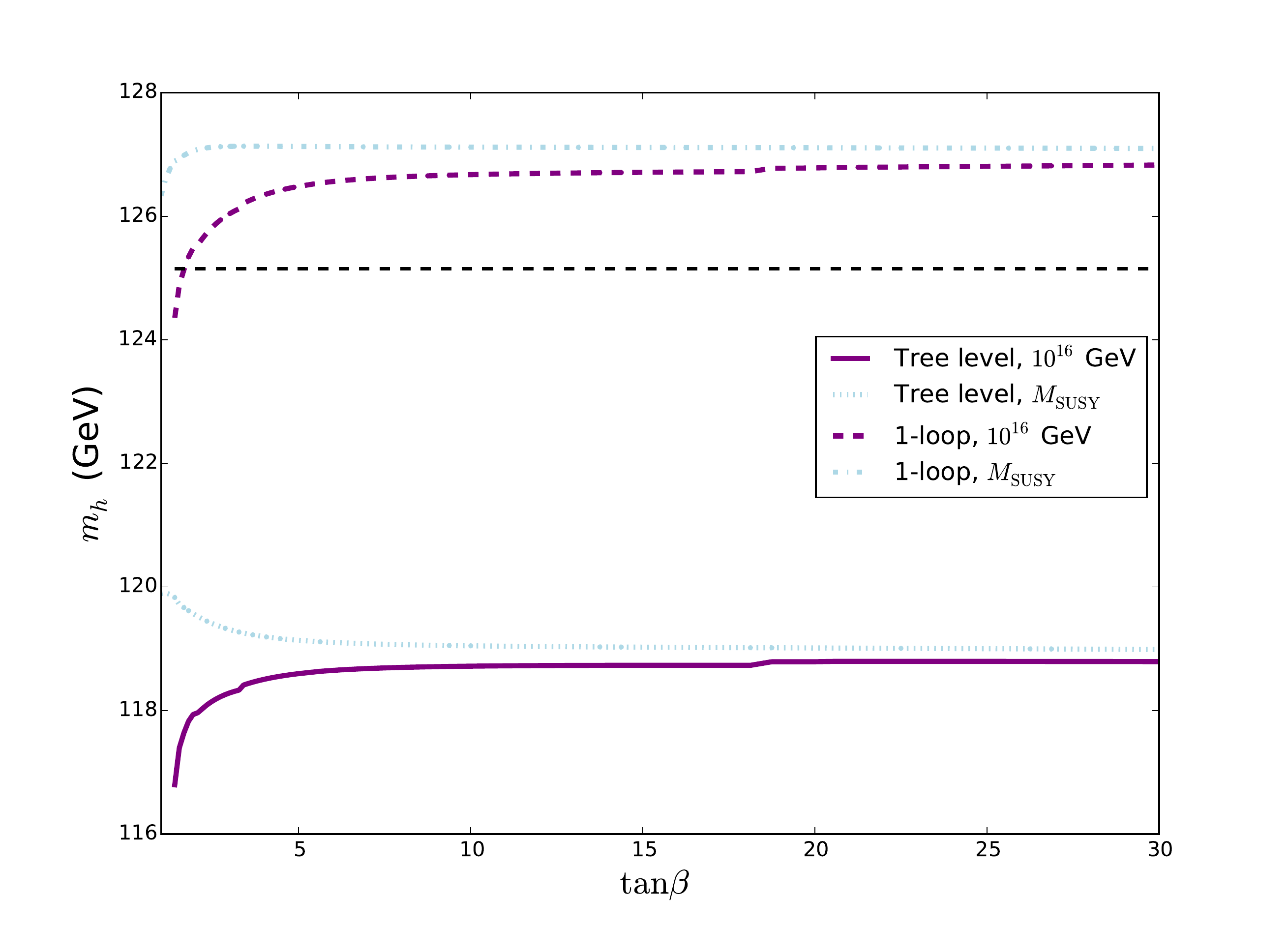}
\caption{Effect of loop corrections in the low-energy theory on the Higgs mass. The tree-level and one-loop values for the Higgs mass are shown against $\tan \beta$ for $N=2$ scales of the stop scale (\MSUSY) and $10^{16}$ GeV; the two-loop value of the Higgs mass is fixed to the black dotted line.}\label{FIG:deltamh}\end{figure}

Finally we consider the effect of the loop corrections in the low-energy theory on the Higgs mass (i.e. those coming from the Higgs sector itself, the top and the electroweakinos). In figure \ref{FIG:deltamh} we show the tree-level and one-loop values for the Higgs mass as we vary $\tan \beta$ (with \MSUSY fixed to ensure $m_h = 125.15$ GeV at two loops). We find a significant upward shift of about $7$ GeV at one-loop, and then a downward shift of about $1$ or $2$ GeV from one to two loops. Note that we can interpret the ``tree-level'' Higgs mass as the loop-level Higgs mass in the full Dirac gaugino model including the effects of the stops and gluinos (which in the EFT formalism appear via the RGEs, rather than fixed-order diagrams).

\begin{figure}[h]\centering
\includegraphics[width=0.8\textwidth]{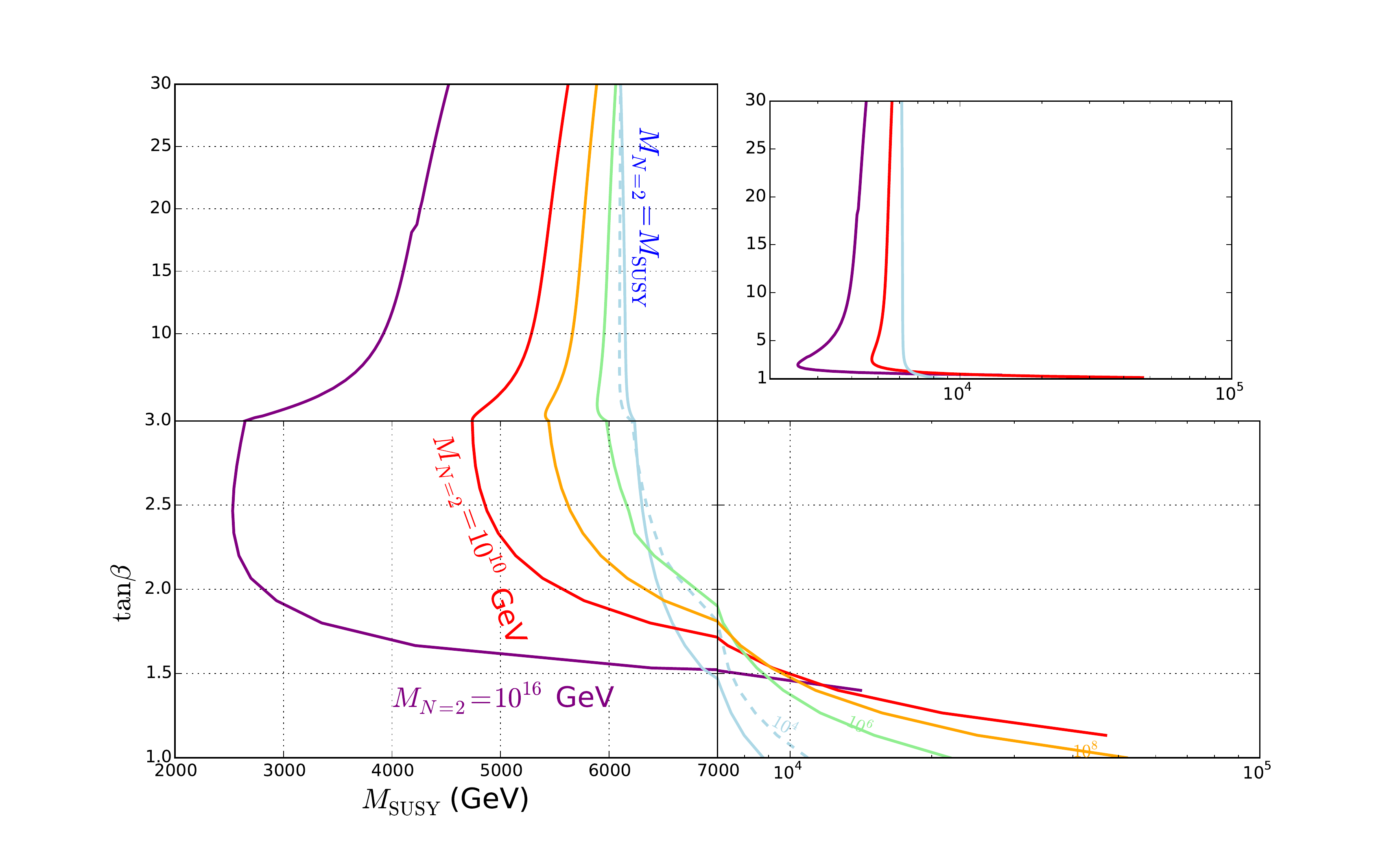}
\caption{SUSY scale that fits $m_h = 125.2$ GeV against $\tan \beta$. The cases $M_{N=2} = \{\MSUSY, 10^{10}\ \GeV, 10^{16}\ \GeV\}$ are the solid lines in blue, red and purple respectively and are labelled in full; the cases $M_{N=2} = \{ 10^{4}, 10^6, 10^8\}$ GeV are respectively shown in blue dashed, solid green and solid orange curves and only labelled with $\{ 10^{4}, 10^6, 10^8\}$. Due to the large range of scales \MSUSY values for small $\tan \beta$ and the little change for large $\tan \beta$ we have split the plot into three quadrants to show the values more clearly, but for comparison we give an inset graph showing the three curves $M_{N=2} = \{ \MSUSY, 10^{10}\ \GeV, 10^{16}\ \GeV\}$ with $\MSUSY$ (GeV) on a logarithmic scale on the abscissa and $\tan \beta$ on a linear scale on the ordinate. }
\label{FIG:mhsusy}\end{figure}

In figure \ref{FIG:mhsusy} we show the final curve of $\tan \beta$ against \MSUSY, for different values of the $N=2$ scale between \MSUSY and $10^{16}$ GeV. 

The plot shows that there is a minimum for \MSUSY around $\tan \beta \simeq 2$ or $3$, particularly for larger values of $M_{N=2}$, which can be understood in terms of the splitting of $\lambda_T$ from its $N=2$ value and the consequent boost to the Higgs mass, which can be clearly seen in figure \ref{FIG:MSUSYZ1}.  

The results in figure \ref{FIG:mhsusy} contrast starkly with the MSSM case matched onto the 2HDM as shown in e.g. \cite{Lee:2015uza}: due to the enhancement to the Higgs mass from the new couplings already seen in figure \ref{FIG:MSUSYZ1} we have a \emph{much} lower SUSY scale. On the other hand, there are significant differences from the values quoted in \cite{Ellis:2016gxa} which are most closely related to the case $M_{N=2} = \MSUSY$; here of course we have light electroweakinos, although the largest difference is the significantly more accurate EFT calculation employed here.

\section{Experimental constraints}
\label{SEC:CONSTRAINTS}

Since our model realises excellent alignment, the light Higgs couplings are very nearly Standard-Model-like across the whole parameter space, and so there is no significant constraint from those -- this is in contrast to e.g. the hMSSM scenario, where for low $\tan \beta$ the Higgs couplings provided until recently the most important lower bound on the Heavy Higgs mass. However, there are still significant constraints on the parameter space coming from electroweak precision tests, flavour and direct searches, as we detail below.

\subsection{Electroweak precision corrections}

There are two contributions to the electroweak precision parameters: those coming from the high-energy theory, and those coming from the low-energy theory. In the high-energy theory there will be contributions at tree-level from the triplet scalars: they should obtain a vacuum expectation value, and in our EFT this manifests itself as generating effective operators. 

In the limit of zero CP violation, and neglecting the terms $A_S, A_T$ we can write the effective operator arising from integrating out the triplet as quite simply
\begin{align}
\lagr \supset& \frac{1}{4m_{TP}^4} \mathrm{tr} \bigg[ D_\mu \bigg( \sigma^a \big[ (\sqrt{2} \lambda_T \mu + g_2 m_{D2} ) H_d^\dagger \sigma^a H_d +  ( g_2 m_{D2} - \sqrt{2} \lambda_T\mu)  H_u^\dagger \sigma^a H_u \big]\bigg) \bigg]^2
\end{align}
where we understand summation on the index $a$ and 
\begin{align}
D_\mu \sigma^a = \sigma^a \partial_\mu -ig_2 [W_\mu, \sigma^a].
\end{align}
When we give a vacuum expectation value to the Higgs, this translates into the constraint from the expectation value of the triplet:
\begin{align}
\Delta \rho =& \frac{\Delta m_W^2}{m_W^2} = \frac{v^2}{m_{TP}^4} \bigg( \sqrt{2} \lambda_T \mu + g_2 m_{D2} c_{2\beta}  \bigg)^2 ,
\end{align}
while the experimental best-fit value is \cite{PDG}
\begin{align}
\Delta \rho = (3.7 \pm  2.3) \times 10^{-4}.
\end{align}
For $\mu = 500$ GeV and an approximately $N=2$ value for $\lambda_T$, with small $\tan \beta$ insisting that this contribution does not exceed the experimental bound by $3\sigma$ gives
\begin{align}
m_{TP} > 1500\ \GeV
\end{align}
while simply saturating without exceeding the central best-fit value would limit instead $m_{TP} > 2\ \TeV $.

\begin{figure}\centering
\includegraphics[width=0.65\textwidth]{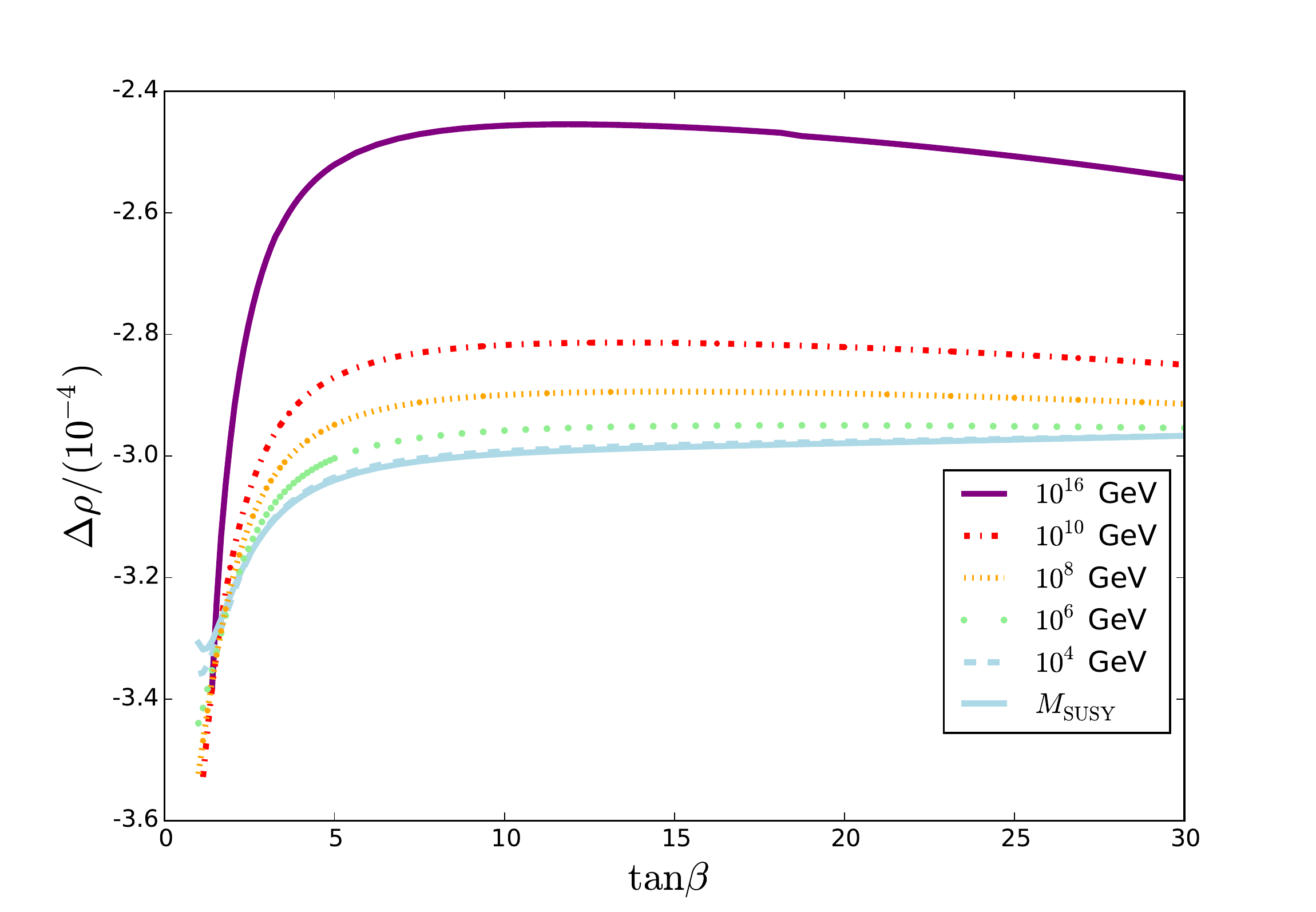}
\caption{$\Delta \rho$ calculated at one-loop in the low-energy theory, for different values of $M_{N=2}$ given in the legend. We see that the magnitude is roughly equal to the experimental error, and we are always well within $3\sigma$ of the experimental central value (which is anyway above the Standard Model value by $1.6 \sigma$).}
\label{FIG:deltarho}\end{figure}

On the other hand, we also have a contribution from the electroweakinos at loop level, which increases as the Dirac mass/$\mu$-term become smaller. Hence they cannot be arbitrarily light. In figure \ref{FIG:deltarho} we plot the value of $\Delta \rho$ calculated in the low-energy theory for the scan values (\ref{EQ:scanvalues}) and find that they are below the experimental limit across the whole parameter space. 

\subsection{Bounds on $\tan \beta$ and $m_A$}

The most stringent constraints on the parameter space of our model come from the searches for $pp \rightarrow H/A \rightarrow \tau \tau$ at the LHC; and the decay $B \rightarrow s \gamma$ determined in \cite{Misiak:2017bgg}, which bounds the charged Higgs mass to be heavier than $580$ GeV independent of the value of $\tan \beta$ (which in turn bounds the mass of the pseudoscalar Higgs to be above around $568$ GeV).

\begin{figure}\centering
\includegraphics[width=0.8\textwidth]{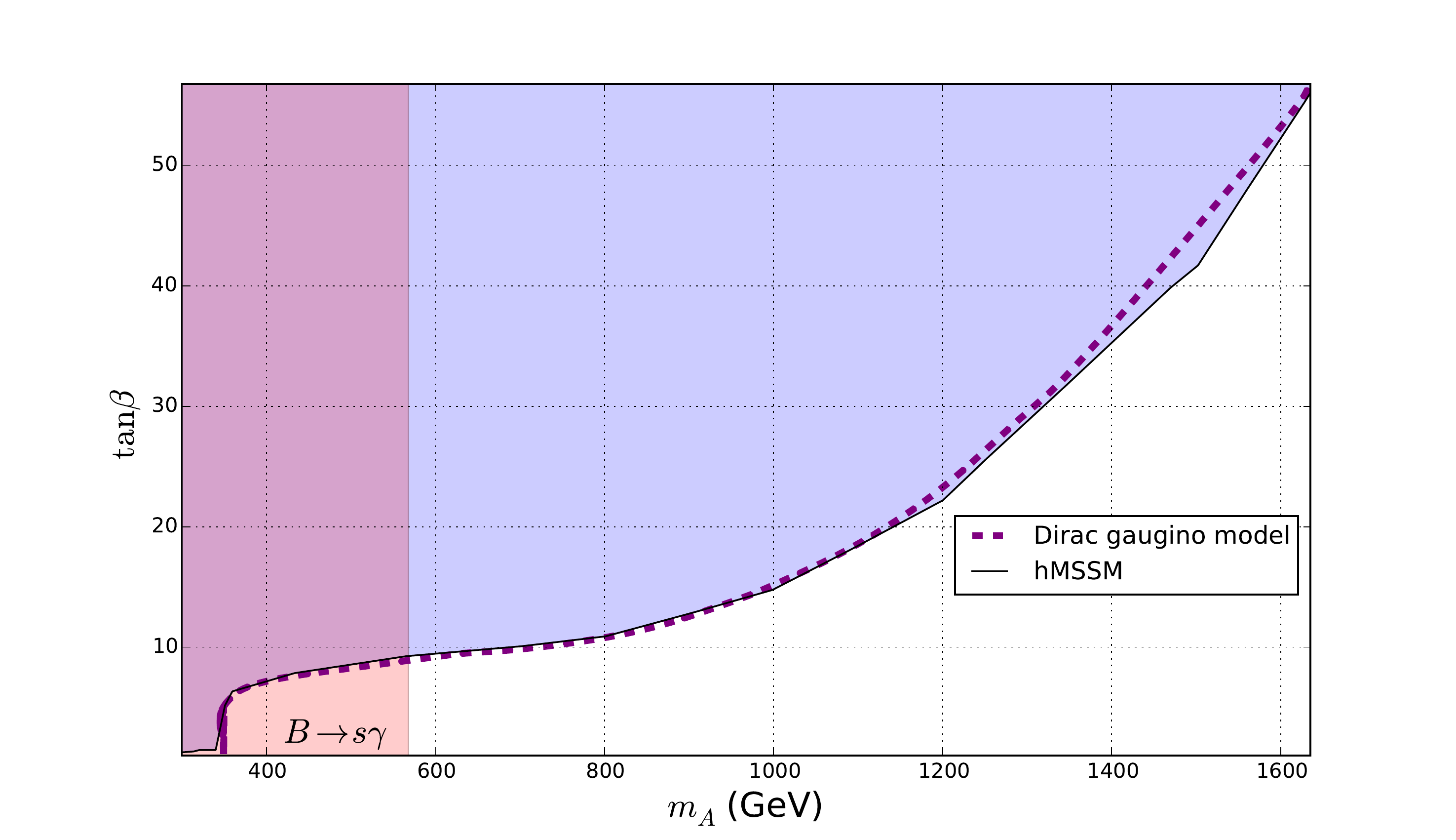}
\caption{Bounds from $pp \rightarrow H/A \rightarrow \tau^+ \tau^-$ (blue region) and $B \rightarrow s \gamma$ (red region, $m_{A} \lesssim 568$ GeV) interpreted in the $m_A/\tan \beta$ plane for the hMSSM (taken from \cite{Aaboud:2017sjh}) and our model. }
\label{FIG:LHCandflavour}\end{figure}

The bounds from run 1 of the LHC were rather mild on the hMSSM: they restricted $\tan \beta < 8$ for low $m_A$ (see e.g. \cite{Carena:2014nza,Aad:2015pla}). In \cite{Ellis:2016gxa} it was claimed that in the h2MSSM these bounds would apply unaltered; while it is true that the couplings to the pseudoscalar are the same in the h2MSSM and hMSSM, the ``heavy'' Higgs does have altered couplings at small $m_A$ and $\tan \beta$ -- since it is more aligned. Since the production of the Heavy Higgs is dominated at small $\tan \beta$ by gluon fusion, and at large $\tan \beta$ by the bbH process, then we would expect some differences at small $\tan \beta$. However, recently, ATLAS produced a much enhanced bound \cite{Aaboud:2017sjh} on gluon fusion and bbH production and then decay to $\tau$ pairs; they also interpreted this in terms of the hMSSM. To compare to our model we computed Higgs production using {\tt SusHi} \cite{Harlander:2012pb,Harlander:2016hcx,Harlander:2002wh,Harlander:2003ai,Actis:2008ug,Harlander:2005rq,Chetyrkin:2000yt} and rescaled the production cross-sections according to b-quark and gluon couplings computed in our {\tt SARAH/SPheno} code, then multiplied by the tau decay branching fraction, and combined the bound assuming that the signals from $H/A$ production overlap for small mass differences. We show the result in figure \ref{FIG:LHCandflavour}, where we also show the bound from \cite{Aaboud:2017sjh} on the hMSSM. We find almost no difference, except that the bound on our model is very slightly weaker once decays to the electroweakinos are permitted. However, the branching ratio to electroweakinos in that region is never significant enough to reduce the $\tau$ decay fraction.

\section{Alignment in the MRSSM}
\label{SEC:MRSSM}

For completeness we now discuss the case of the MRSSM in the same limit as for the DG-MSSM. Since the tree-level THDM parameters are the same as those of the MSSM in the limit of large adjoint scalar and R-Higgs masses,\footnote{So therefore $Z_1, Z_5, Z_6$ are the same as in the MSSM case, i.e. equations (\ref{Z_1_Specific}) - (\ref{Z_6_Specific}) with $\lambda_S = \lambda_T = 0$.} there is no contribution to $Z_6$ from the running of the parameters $\lambda_{S_{u,d}}, \lambda_{T_{u,d}}$. We can first write the neutral Higgs mass matrix as
\begin{align}
m_{h,H}^2 =\twomat[M_Z^2 c_{2\beta}^2 + v^2 \Delta Z_1, -M_Z^2 s_{2\beta} c_{2\beta} + v^2 \Delta Z_6 ][-M_Z^2 s_{2\beta} c_{2\beta} + v^2 \Delta Z_6 , m_A^2 + M_Z^2 s_{2\beta}^2  + v^2 \Delta Z_5].
\end{align}
If we consider the loop corrections due to $\lambda_{S_{u,d}}, \lambda_{T_{u,d}}$ to be small, then the analysis of alignment is identical to the MSSM case, and we can apply the hMSSM logic. 
However, if we instead take them to be non-negligible -- such as in \cite{Diessner:2014ksa,Bertuzzo:2014bwa,Diessner:2015yna,Diessner:2015iln} -- then the contributions to $\lambda_2$ no longer dominate, and the hMSSM reasoning may no longer apply. On the other hand, the largest contribution from the other particles will still be to $\lambda_2$, and so we can assume that
\begin{align}
\lambda_2 =& \frac{M_Z^2 + \epsilon}{v^2} , \qquad \lambda_1 = \frac{M_Z^2}{v^2} + \delta \lambda_1\nn\\
\lambda_5 =& 0, \qquad \lambda_{345} = \lambda_{34} = - \frac{M_Z^2}{v^2} + \delta \lambda_{34}.
\end{align}
To eliminate $\epsilon$, we eliminate $\lambda_2$ in terms of the Higgs mass, which for general $\lambda_i, i=1...4$ (and $\lambda_5 = \lambda_6 = \lambda_7 =0$):
\begin{align}
\lambda_2 v^2 s_\beta^2 + m_A^2 c_\beta^2 =& m_h^2 + \frac{s_\beta^2 c_\beta^2 ( m_A^2 -\lambda_{34} v^2 )^2}{\lambda_1 v^2 c_\beta^2 + m_{A}^2 s_\beta^2 - m_h^2};
\label{EQ:hTHDM}\end{align}
this can then be substituted into the expression for $Z_6$: 
\begin{align}
Z_6 =& - \frac{  s_\beta c_\beta}{\lambda_1 v^4 c_\beta^2 + m_{A}^2 v^2 s_\beta^2 - m_h^2 v^2} \times \nn\\
& \bigg[ (\lambda_1 v^2 c_\beta^2 - m_h^2)  (\lambda_1 v^2 c_\beta^2 - m_h^2 + m_A^2 - \lambda_{34} v^2 c_{2\beta}) + \lambda_{34} s_\beta^2 v^2 (m_A^2  -  \lambda_{34} c_\beta^2  v^2) \bigg].
\label{EQ:hTHDMZ6}\end{align}

We give the loop corrections to the $\lambda_i$ from the adjoint scalars in appendix \ref{APP:MRSSM}, but in the simplified case of $m_{T^+} = m_{T^-} = m_{SR} = m_{SI} = M_{\Sigma}$ and $g_Y = g_2 =0$ we have, for matching at a scale $\mu$:
\begin{align}
 \delta \lambda_1 =& \frac{1}{16\pi^2} \log \frac{M_{\Sigma}^2}{\mu^2} \bigg(5 \lambda_{T_d}^4  + 2 \lambda_{S_d}^2 \lambda_{T_d}^2  + \lambda_{S_d}^4\bigg)  \nn\\
 \delta\lambda_2 =& \frac{1}{16\pi^2} \log \frac{M_{\Sigma}^2}{\mu^2} \bigg( 5 \lambda_{T_u}^4 + 2 \lambda_{S_u}^2 \lambda_{T_u}^2  + \lambda_{S_u}^4  \bigg) \nn\\
  \delta\lambda_3 =& \frac{1}{16\pi^2} \log \frac{M_{\Sigma}^2}{\mu^2} \bigg( 5 \lambda_{T_d}^2 \lambda_{T_u}^2 + \lambda_{S_d} \lambda_{S_u} \lambda_{T_d} \lambda_{T_u}  + \lambda_{S_d}^2 \lambda_{S_u}^2 \bigg)\nn\\
 \delta\lambda_4 =& \frac{1}{16\pi^2} \log \frac{M_{\Sigma}^2}{\mu^2} \bigg(- 4 \lambda_{T_d}^2 \lambda_{T_u}^2 - 4 \lambda_{T_d} \lambda_{T_u} \lambda_{S_d} \lambda_{S_u}  \bigg).
\end{align}
If we then take (as in  \cite{Diessner:2014ksa,Diessner:2015yna,Diessner:2015iln}) $\lambda_{S_u} = -\lambda_{S_d} \equiv \lambda, \lambda_{T_u} =  \lambda_{T_d} \equiv \Lambda$, and allow an additional contribution $\epsilon/v^2$ to $\lambda_2$ from the stops,  then we have
\begin{align}
 Z_6 =& - \frac{1}{2} s_{2\beta} c_{2\beta} \bigg(\frac{2M_Z^2}{v^2} +  \frac{2 \Lambda^4 }{16\pi^2}\log \frac{M_{\Sigma}^2}{\mu^2} \bigg)+ \frac{\epsilon}{v^2} s_\beta^3 c_\beta \nn\\
\Delta Z_1 =&  \frac{1 }{16\pi^2}\log \frac{M_{\Sigma}^2}{\mu^2}\bigg[\lambda^4 + 2 \lambda^2 \Lambda^2 + 3 \Lambda^4 + 2 \Lambda^4 c_{2\beta}^2\bigg] + \frac{\epsilon}{v^2} s_\beta^4.
\end{align}
We see that when the couplings $\lambda, \Lambda$ are large enough, the alignment will always be improved compared to the MSSM, because the enhancement to $Z_1$ is always greater than that to $Z_6$. We note three cases of particular interest:
\begin{enumerate}
\item If we increase the contributions from the adjoint scalars to the point that we can neglect those from the stops, then we see that for small $\tan \beta$ we will easily have alignment (in contrast to the MSSM case). 
\item Alternatively, we could enhance the contributions from $\lambda$ rather than $\Lambda$, since the former coupling does not contribute to $Z_6$. 
\item On the other hand, if we take the $N=2$ supersymmetric limit 
\begin{align}
\lambda_{T_u} =& \lambda_{T_d} = \frac{g_2}{\sqrt{2}},\quad \lambda_{S_u} = \frac{g_Y}{\sqrt{2}}, \quad\lambda_{S_d} =-\frac{g_Y}{\sqrt{2}},
\label{EQ:MRSSMNeq2}\end{align}
we find, using the expressions in appendix \ref{APP:MRSSM} (and no longer neglecting the gauge couplings):
\begin{align}
 \delta \lambda_1 =& \frac{1}{16\pi^2} \log \frac{M_{\Sigma}^2}{\mu^2} \frac{1}{4}\bigg( 3 g_2^4 + 2 g_2^2 g_Y^2 + g_Y^4 \bigg)  \nn\\
 \delta\lambda_2 =& \delta \lambda_1 \nn\\
  \delta\lambda_3 =&\frac{1}{16\pi^2} \log \frac{M_{\Sigma}^2}{\mu^2} \frac{1}{4}\bigg( 3 g_2^4 - 2 g_2^2 g_Y^2 + g_Y^4 \bigg) \nn\\
 \delta\lambda_4 =&  \frac{1}{16\pi^2} \log \frac{M_{\Sigma}^2}{\mu^2} \bigg( g_2^2 g_Y^2\bigg)
\end{align}
giving 
\begin{align}
\delta \lambda_{345} =& \delta \lambda_1 \equiv \delta \lambda,
\end{align}
so there is no shift to $Z_6$ from the adjoint scalars, but we do have a shift to $Z_1$, i.e
\begin{align}
\Delta Z_1 =& \delta \lambda , \qquad \Delta Z_6 =0.
\end{align}
If the mass of the adjoint scalars is comparable to the mass of the stops, then this will however never be significant. On the other hand, if we take the adjoint scalars to be very heavy, then this indicates that we can have improved alignment relative to the MSSM. To quantify this, we can use our above expression for $Z_6$ (\ref{EQ:hTHDMZ6}): 
\begin{align}
  Z_6 = &  - \frac{  s_\beta c_\beta}{(M_Z^2 v^2 + \delta \lambda v^4) c_\beta^2 + m_{A}^2 v^2 s_\beta^2 - m_h^2 v^2}  \bigg[ \Delta_0 + \delta \lambda v^2 ( m_A^2 - m_h^2 + 2c_\beta^2 M_Z^2) \bigg] 
\end{align}
where
\begin{align}
\Delta_0 =& m_h^2 ( m_h^2 - m_A^2 - M_Z^2 (4 c_\beta^2 - 1) ) +M_Z^2c_{2\beta} ( m_A^2 +2 M_Z^2 c_\beta^2 )  
\end{align}
which is the numerator for the MSSM case. In the case that $m_A^2 \gg m_h^2$ (which corresponds to our case of interest -- even though we would like $m_A$ small enough to not entirely be in the decoupling limit), we therefore find
\begin{align}
Z_6 \simeq& \frac{  1 }{ t_\beta} \bigg[ \frac{m_h^2 - M_Z^2 c_{2\beta}}{v^2} - \delta \lambda  \bigg].
\end{align}
For $M_\Sigma = 100 M_{SUSY}$ (a rather extreme value) and matching at $M_{SUSY}$ we therefore find 
\begin{align}
\delta \lambda \simeq & 0.04 \,\frac{m_h^2}{v^2},
\end{align}
and so the deviation of $Z_6$ from the MSSM value due to the adjoint scalars should be less than $4\%$. On the other hand, as we shall see below, they can still have a significant effect on the SUSY scale.
\end{enumerate}

Therefore, from the analysis above, in all three cases of interest, the alignment will never be as good as for our minimal Dirac gaugino model, because of the tree-level contribution to misalignment: we shall illustrate this for the $N=2$ case in the next subsection.

\subsection{Numerical analysis of an $N=2$ MRSSM}

To compare with our previous analysis of the DG-MSSM, here we present a simplified numerical analysis for an $N=2$ MRSSM, as defined in point 3 above and equation (\ref{EQ:MRSSMNeq2}). From our estimations above, the alignment should only differ from the MSSM when relatively extreme values are taken for the adjoint scalar masses, and so to perform a precise analysis we would need to have a tower of effective field theories and the appropriate threshold corrections. Instead we decided to neglect all loop-level threshold corrections other than those from the adjoint scalars (although we use 2-loop RGEs throughout) and performed a simple analysis where the low energy model was approximated by the Standard Model and type-II two-Higgs doublet model. In this way we should obtain an idea of how the adjoint scalar masses cause the SUSY scale and alignment to vary from the predictions of the MSSM. 

\subsubsection{Procedure}

Two-loop Standard Model matching values were implemented at $m_{t}$ for the standard model gauge, Yukawa, and Higgs quartic couplings from \cite{Buttazzo:2013uya} and a two-loop Standard Model running was performed up to an intermediate scale $Q=600$ GeV, where the $\lambda_i(Q)$ couplings were given approximate values to be determined through future iterations between the scales $Q$ and $M_{N=2}$. The two-Higgs doublet model 2-loop running was implemented up to the supersymmetry breaking scale defining the leading squark masses, $M_{SUSY}$, where guesses were made for the inputs of the parameters $\lambda_{S,T_{u,d}}$. The MRSSM was then run to 2-loops to some high scale $M_{N=2}$ where the $N=2$ boundary conditions \eqref{EQ:MRSSMNeq2} were implemented. All two-loop beta functions were generated in $\tt{SARAH}$, and the value of $m_A^{tree} = 600$ GeV was taken as in the minimal Dirac gaugino case. In this simplified model, as the electroweakinos are not taken to be light, the intermediate scale $Q$ is taken to match the choice of heavy Higgs mass. Indeed, with these choices we should understand the Dirac gaugino masses $m_{DY}, m_{D2}, m_{D3}$ and the higgsino mass to be at \MSUSY, and also the masses of the R-Higgs fields $R_{u,d}$ should be at that scale, because we do not implement any threshold corrections from those fields (leaving these to future work). 

On the run down, $\lambda_i(M_{SUSY})$ were matched to the 1-loop threshold corrections coming from the heavy $S,T$ scalars as given in appendix \ref{APP:MRSSM}, taking the adjoint scalars to be degenerate with mass $M_{\Sigma}$. This process was iterated, re-matching the gauge and Yukawa couplings onto their 2-loop Standard Model running values at the scale $Q$, while the $\lambda_i$ and $\lambda_{S,T_{u,d}}$ couplings were matched to the outputs from the previous running until their values converged. Finally, the $\lambda_i$ parameters were mapped back onto the Higgs quartic coupling using $\lambda(Q) = Z_1(Q)$ and the Standard Model couplings were run back down to $m_{t}$. The correct Higgs mass was selected from the criterion $\lambda(m_{t}) = 0.252 \pm 0.002$, corresponding to a pole Higgs mass of $m_h = 125 \pm 0.5$ GeV.

This process was executed for scans over the values $\tan\beta \in [2,20]$; $M_{SUSY} \in [0.5,10]$ TeV; $M_{\Sigma} = \left\{ 5, 10, 100\right\} M_{SUSY}$ and $M_{N=2} = \left\{10^6, 10^{10}, 10^{16}\right\}$ GeV.

\subsubsection{Running from the N=2 scale to $M_{SUSY}$}


\begin{figure}[H]\centering
\includegraphics[width=0.8\textwidth]{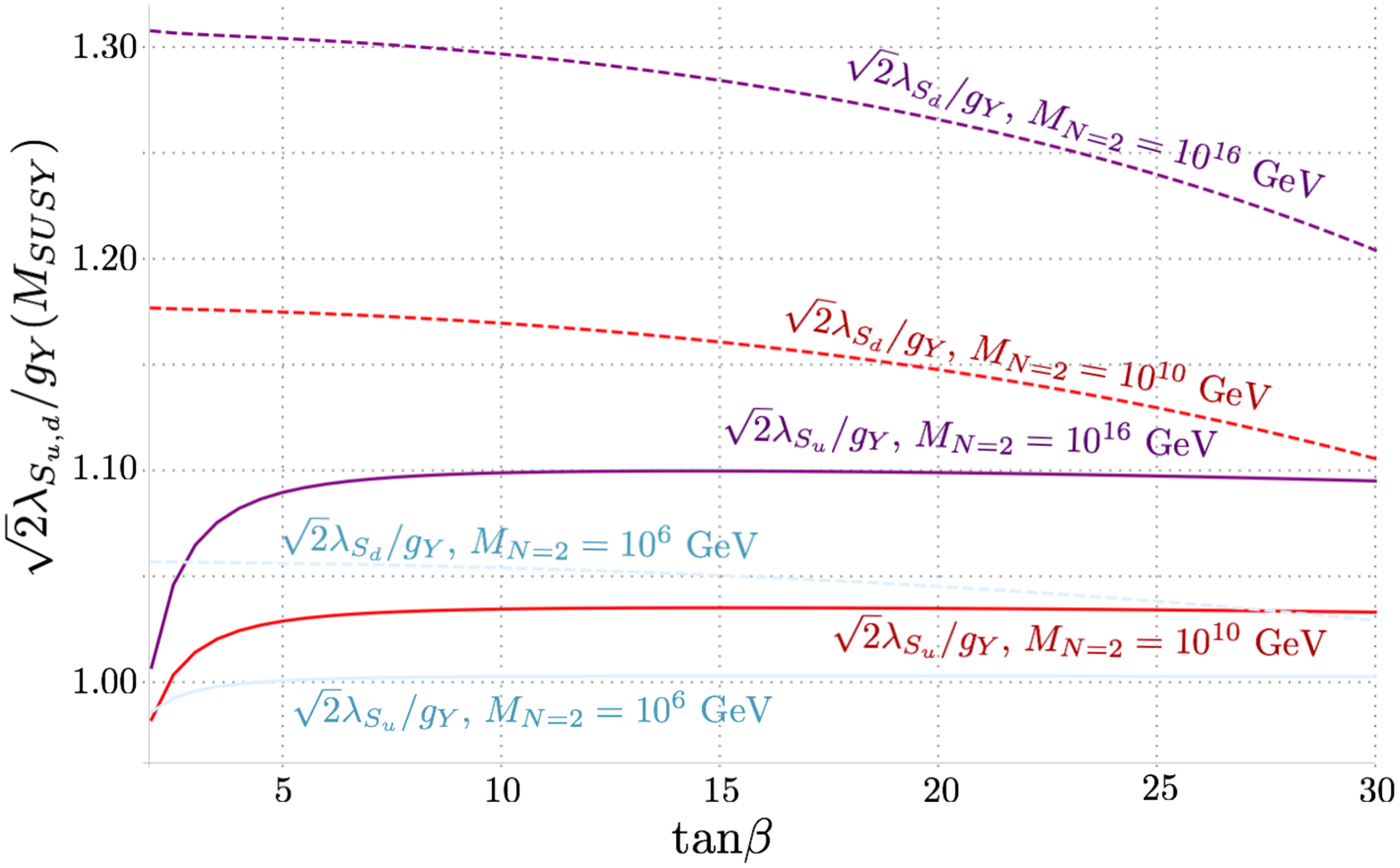}
\vspace{-35pt}
\caption{Variations in the ratio $\sqrt{2} \lambda_{S_{u,d}}/g_{Y}$ against $\tan \beta$ at the $M_{SUSY}$ scale for $N=2$ scales $10^6, 10^{10}$ and $10^{16}$ GeV. }
\label{S_V_Tan}
\end{figure}

\begin{figure}[h]
\begin{center}
\includegraphics[width=0.8\textwidth]{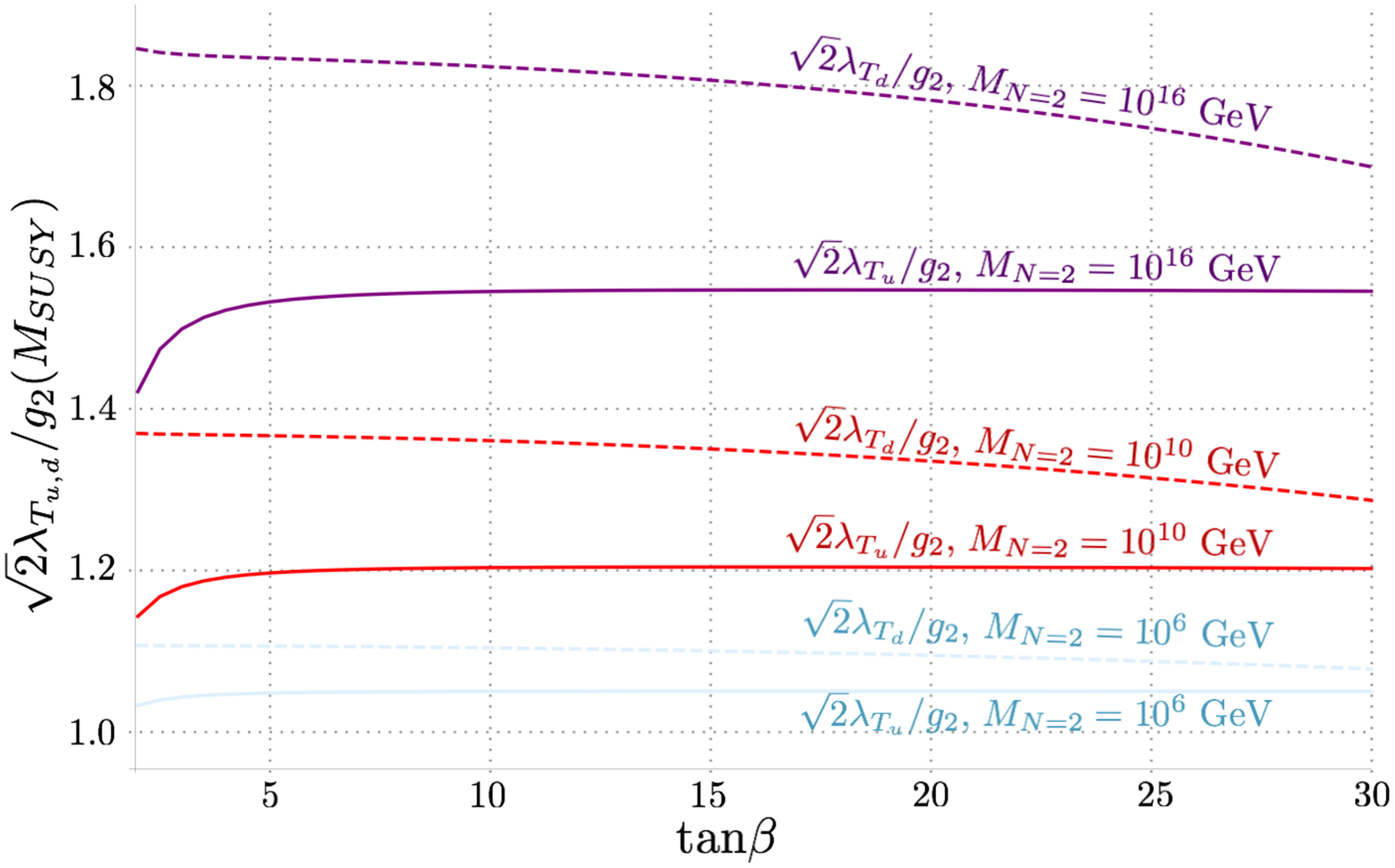}
\end{center}
\vspace{-35pt}
\caption{Variations in the ratio $\sqrt{2} \lambda_{T_{u,d}}/g$ against $\tan \beta$ at the $M_{SUSY}$ scale for $N=2$ scales $10^6, 10^{10}$ and $10^{16}$ GeV.}
\label{T_V_Tan}
\end{figure}

The ratios in figures (\ref{S_V_Tan}, \ref{T_V_Tan}) are taken with a common $M_{SUSY}$ scale of $10$ TeV, while the associated value of $m_h$ is unconstrained. $M_{\Sigma}$ is kept fixed - in figures (\ref{S_V_Tan}, \ref{T_V_Tan}) chosen as $M_{\Sigma} = 10 \, M_{SUSY}$. Here the modulus of the ratio is plotted, since the $\lambda_{S_d}$ ratio is negative to respect the $N=2$ supersymmetry relations.  
 As expected, the model is closest to the alignment limit when the $N=2$ scale is closer to the $M_{SUSY}$ scale. 
It can be seen that the Higgs mass is boosted to a greater extent by the down-type couplings than the up-type, where the ratio $\sqrt{2} \lambda_{T_d}/g$ has the largest effect, especially for higher values of $N=2$ scale.

\begin{figure}[H]
\begin{center}
\includegraphics[width=0.8\textwidth]{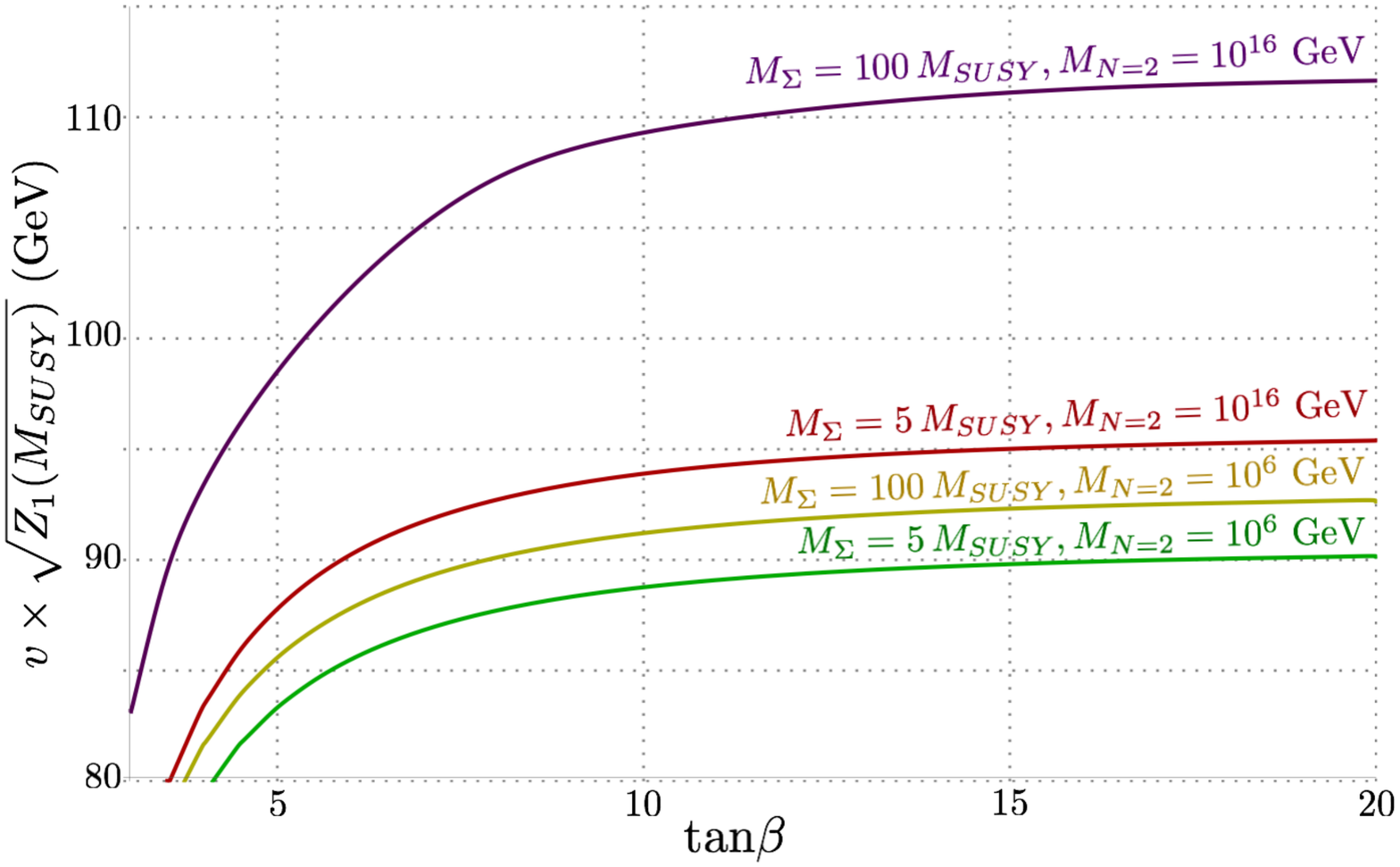}
\vspace{-35pt}
\end{center}
\caption{$v \sqrt{Z_1(M_{SUSY})}$ against $\tan \beta$ for $N=2$ scales $10^6$ and $10^{16}$ GeV, corresponding to the ``tree-level'' value of $m_h$ before running down in low-energy effective theory.}
\label{Z1_MRSSM}
\end{figure}

Figure \ref{Z1_MRSSM} shows the ``tree-level'' Higgs mass against $\tan\beta$ before running down from $M_{SUSY}$ (where the value of the Higgs mass calculated at $m_t$ matches the experimental value). For the lowest values of $M_{N=2}$ and $M_{\Sigma}$ plotted, $v \sqrt{Z_1(M_{SUSY})}$ is approximately $M_Z$, and where the former increase, so does the boost to the Higgs mass. This boost grows substantially for the simultaneously highest values of $M_{N=2}$ and $M_{\Sigma}$, owing to the large (almost non-perturbative) $\lambda_T$ couplings. While not shown here, it should be noted that even for $M_{N=2} = 10^{10}$ GeV and $M_{\Sigma} = 100 \, M_{SUSY}$, $v \sqrt{Z_1(M_{SUSY})}$ replicates almost identical behaviour to the red curve for  $M_{N=2} = 10^{16}$ GeV and $M_{\Sigma} = 5 \, M_{SUSY}$ shown here.

\subsubsection{Running from $M_{SUSY} \rightarrow Q \rightarrow m_{t}$}

\begin{figure}[h]
\begin{center}
\includegraphics[width=0.8\linewidth]{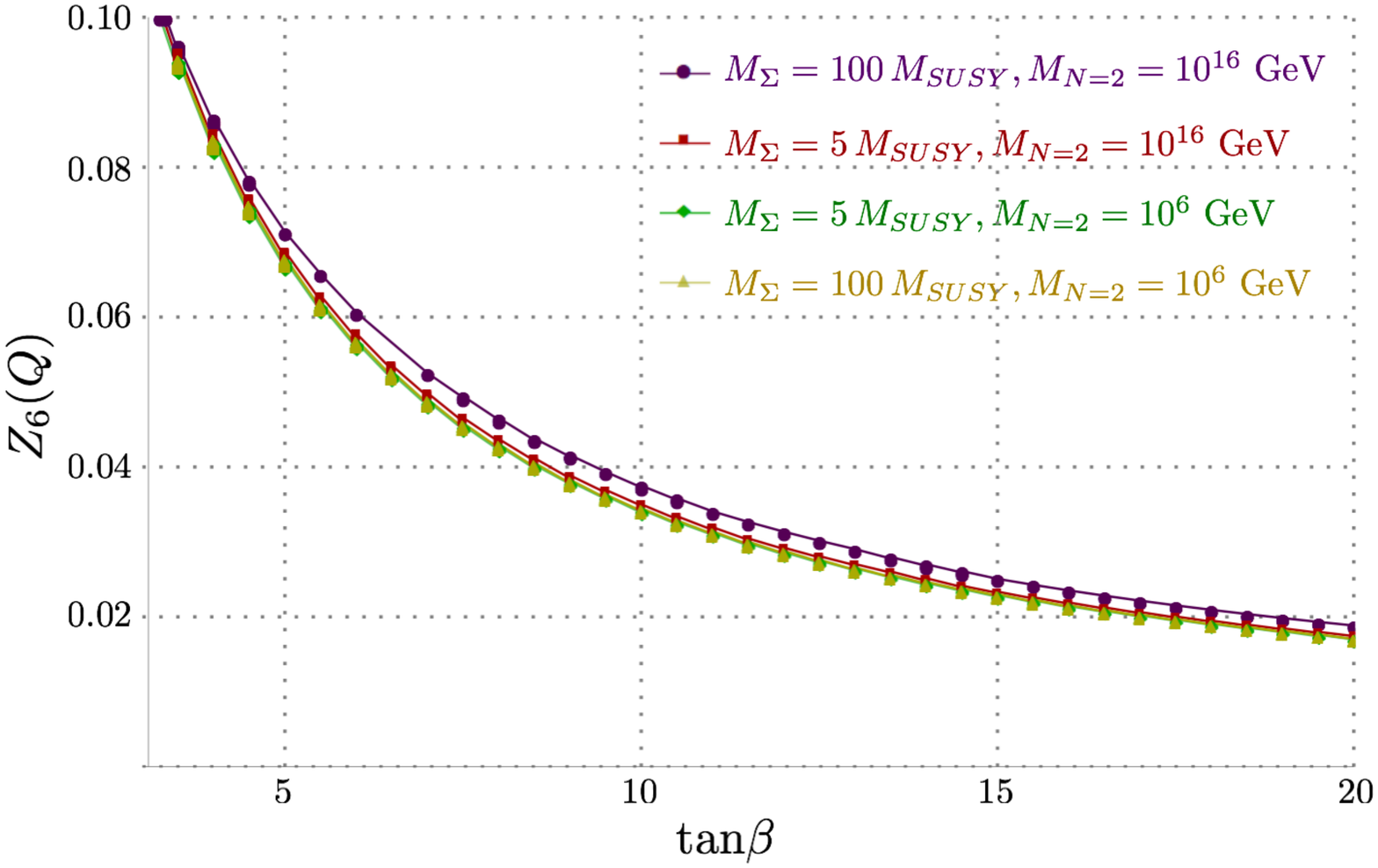}
\end{center}
\vspace{-35pt}
\caption{$Z_6(Q)$ against $\tan\beta$ for a Higgs mass of $125$ GeV at $m_{t}$, for values of $M_{\Sigma}$ = 5, 100 $M_{SUSY}$ and $M_{N=2} = 10^6$ and $10^{16}$ GeV.}
\label{Z6_MRSSM}
\end{figure}

\begin{figure}[H]
\begin{center}
\includegraphics[width=0.8\linewidth]{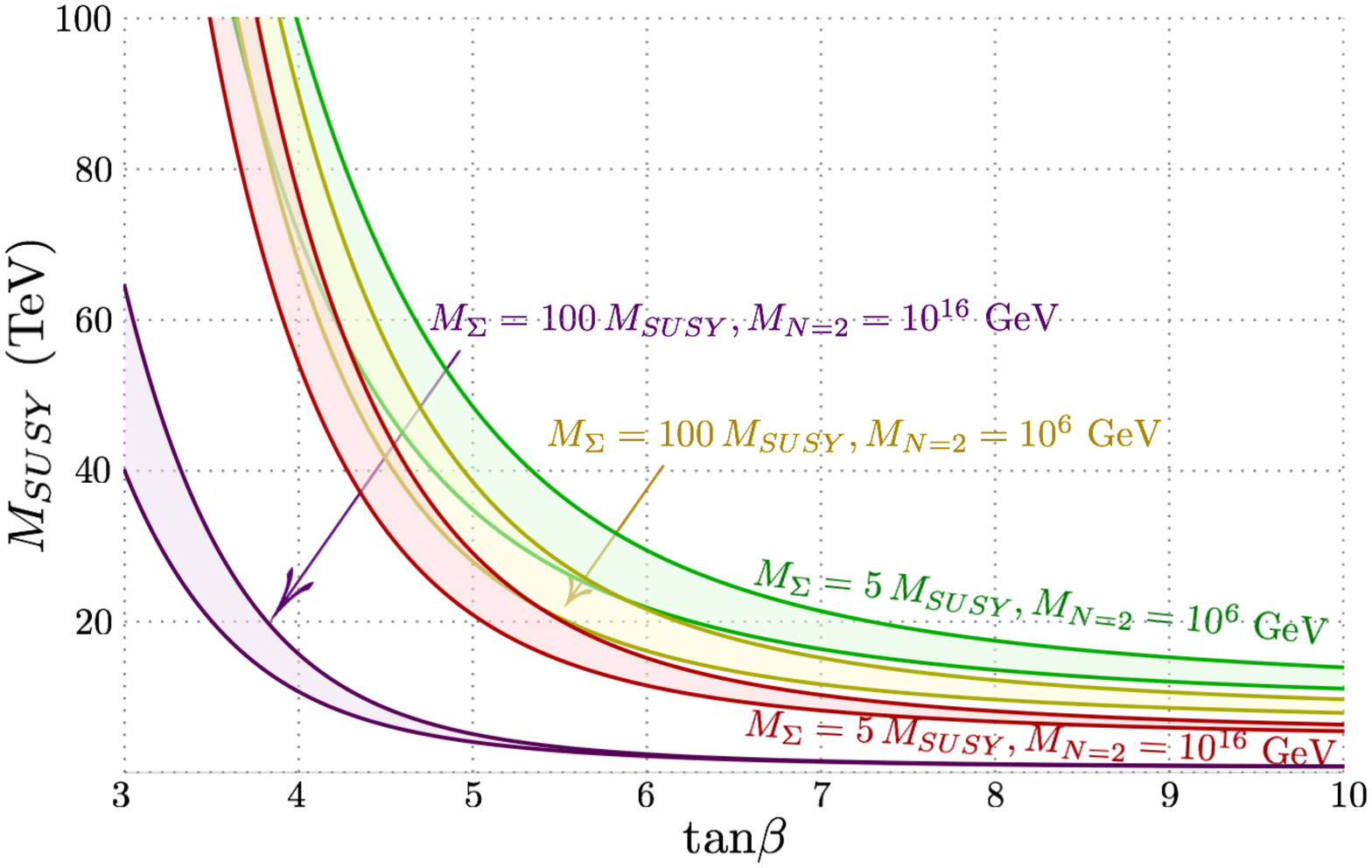}
\end{center}
\vspace{-35pt}
\caption{$M_{SUSY}$ against $\tan\beta$ for a Higgs mass of $125$ GeV at $m_{t}$ and where $Q= 600$ GeV, plotted for values of $M_{\Sigma}$ = 5, 100 $M_{SUSY}$ and $M_{N=2} = 10^6$ and $10^{16}$ GeV.}
\label{Tan_V_Msusy}
\end{figure}

Figure \ref{Z6_MRSSM} shows little deviation in the results for $Z_6(Q)$, regardless of $M_{\Sigma}$ and $M_{N=2}$. Indeed, as anticipated above, the results are almost indistinguishable from the MSSM case, since the adjoint scalars in the MRSSM never give a large boost to the quartic couplings even for the extreme cases we have taken. Exceptionally, the couplings in the case of very heavy scalars and very high $M_{N=2}$ are considerably enhanced and deviate from the $N=2$ relations, making the alignment in this case just marginally worse. While the adjoint scalars give only a very small boost to the Higgs mass, on the other hand it is enough to cause noticeable effects in the predicted $M_{SUSY}$ scale, shown in figure \ref{Tan_V_Msusy}, because of the logarithmic nature of the contributions from other SUSY states.

Figure \ref{Tan_V_Msusy} shows the values of $M_{SUSY}$ against $\tan \beta$ over the parameter scan producing a Higgs mass corresponding to $m_h = 125 \pm 0.5$ GeV: this margin is reflected in the enclosed transparent area\footnote{The variation of $0.5$ GeV is, however, not to imply the total error (which is hard to estimate, but should be comparable to this value although it may be smaller for large values of $M_{SUSY}$) but to give an indication of the sensitivity of the results on the final Higgs mass value.}. For $\tan \beta < 4$, $M_{SUSY}$ is required to be, at the very least, 20 TeV for the highest values of $M_{N=2}$ and very heavy $M_{\Sigma}$, and is closer to $\sim 100$ TeV for lower values. $M_{SUSY}$ stabilises around $\tan \beta = 10$ for all values of $M_{\Sigma}$ and $M_{N=2}$, where at this point $M_{SUSY}$ can be as low as several hundred GeV for $M_{N=2} = 10^{16}$ and very heavy scalars. In this final extreme case (which is of course excluded experimentally, but we give as an indication of the possible effects) the logarithms being resummed in the RGEs become smaller, it is possible that any neglected threshold effects could make a significant difference and the results become unreliable, but we leave this additional analysis to future work.

\section{Conclusions}
\label{SEC:CONCLUSIONS}

We have considered the the consequences for the simplest realisation of Dirac gaugino models when we impose $N=2$ supersymmetric boundary conditions for the Higgs/gauge sector at some energy scale. We found that the model naturally realises alignment in the Higgs mass matrix, and that surprisingly this is preserved even by quantum corrections. Even more interestingly, the departure from $N=2$ relations due to running of the couplings actually leads to both an enhanced Higgs mass (and thus lower SUSY scale/more natural model) and also \emph{improved} alignment when we take the effects of the squarks into account. 

We have provided the most accurate calculation to date for the SUSY scale for a Dirac gaugino model by employing the effective field theory approach, with one-loop boundary conditions at the high scale and two loops at the THDM scale. This leads to the prediction that the scale of coloured superpartners should be above $3$ TeV (when we allow a very high scale for the breaking of the approximate $N=2$ SUSY) but across most of the parameter space it is below $10$ TeV. While this is not encouraging for the detection of stops/gluinos at the LHC, this is well within the reach of a future $100$ TeV collider. On the other hand, the LHC or a future $e^+ e^-$ collider should be able to explore the electroweak sector of the model, including the Higgs sector and the electroweakinos (if they are light).

There are many possible avenues for future work: improving the accuracy of the matching at $\MSUSY$ (as noted recently, matching at two-loop order is often necessary for accuracy of the loop expansion to include all non-logarithmic corrections \cite{Braathen:2017jvs}, although in this class of models as we have discussed all of the missing corrections are believed to be small) and including the effects of the electroweakinos in the matching at one loop, so that we can consider the model with $m_{DY} \sim m_{D2} \sim \MSUSY$; also with the full set of thresholds we could perform an estimate of the error in the calculation (which, again, should already be small -- see e.g. the estimates for the MSSM case in \cite{Vega:2015fna}); or including the effects of possible R-symmetry violating terms. On the other hand, it would also be interesting to more fully explore the consequences for different Dirac gaugino models, such as the MRSSM, where we have only performed a preliminary analysis.


\vskip.3in
\noindent
\section*{Acknowledgments}

 We are grateful to  P.~Slavich for many helpful discussions and comments on the manuscript. S.~W. thanks J.~Harz for help at the very early stage of the project.  We acknowledge the support of the European Research Council (ERC) under the Advanced  Grant Higgs@LHC (ERC-2012-ADG20120216-321133) and the Agence Nationale de Recherche under grant ANR-15-CE31-0002 ``HiggsAutomator''. This work is also supported by the Labex ``Institut Lagrange de Paris'' (ANR-11-IDEX-0004-02,  ANR-10-LABX-63)).

\noindent

\appendix

\section{THDM with light electroweakinos}
\label{APP:THDMEW}

The limit that we are interested in has the electroweakinos much lighter than the singlet and triplet scalars; in order to avoid washing out the tree-level Higgs quartic coupling and generating a large contribution to $\rho$ they should be light. At energies below the stop/sbottom masses, then, we have an effective theory of the two-Higgs doublet model augmented by light electroweakinos. This looks a little like Split supersymmetry or the scenario of \cite{Bagnaschi:2015pwa} (which considered a split scenario with both Higgs doublets light), except that our electroweakinos have Dirac masses and our gluino is heavy; here are therefore new Yukawa couplings between the Higgs doublets $\Phi_i$, the left and right bino $\tilde{B}_i$ and wino $\tilde{W}^a_i$ for $i=1,2$, and the higgsinos $\tilde{h}_{u,d}$:
\begin{align}
\lagr \supset - \frac{1}{\sqrt{2}} \bigg[\tilde{g}_{1u}^{ij} \Phi_i^* \tilde{B}_j \tilde{h}_u +  \tilde{g}_{2u}^{ij} \Phi_i^* \tilde{W}_j^a \sigma^a \tilde{h}_u +\tilde{g}_{1d}^{ij} \Phi_i \tilde{B}_j \tilde{h}_d + \tilde{g}_{2d}^{ij} \Phi_i \tilde{W}_j^a \sigma^a \tilde{h}_d + h.c. \bigg].
\end{align}
This gives neutral and charged fermion mass matrices 
\begin{align}
&\mathcal{M}_{\chi^0} = \left( 
\begin{array}{cccc} 
 M_B & 0 & - \frac{1}{2} v_k g_{1d}^{ki} &  \frac{1}{2} v_k g_{1u}^{ki} \\
0 & M_W &  - \frac{1}{2} v_k g_{2d}^{ki} & - \frac{1}{2} v_k g_{2u}^{ki} \\
- \frac{1}{2} v_k g_{1d}^{ki} & - \frac{1}{2} v_k g_{2d}^{ki} & 0 & -\mu \\
\frac{1}{2} v_k g_{1u}^{ki}& - \frac{1}{2} v_k g_{2u}^{ki} & -\mu & 0 
\end{array}
\right), \qquad \mathcal{M}_{\chi^\pm} = \left( 
\begin{array}{cc} 
M_W &  \frac{1}{\sqrt{2}} v_k g_{2u}^{ki} \\
 - \frac{1}{\sqrt{2}} v_k g_{2d}^{ki} & \mu
\end{array}
\right)
\end{align}
where the bases are $\chi^0 = (\tilde{B}_1, \tilde{B}_2, \tilde{W}^0_1, \tilde{W}^0_2,\tilde{h}_d^0, \tilde{h}_u^0) $ and the charged mass terms are $\lagr \supset -(\tilde{W}^-_i, h_d^-) \mathcal{M}_{\chi^\pm} \twv[\tilde{W}^+_i, h_u^+]$. Note that $M_B, M_W$ are $2\times 2$ matrices.

At the SUSY scale, we match the above to the corresponding couplings in the Dirac gaugino theory:
\begin{align}
\lagr^{DG} \supset& - \frac{g_Y}{\sqrt{2}} H_u^* \tilde{B}_1 \tilde{h}_u - \frac{g_2}{\sqrt{2}} H_u^* \tilde{W}_1^a \sigma^a \tilde{h}_u + \frac{g_Y}{\sqrt{2}} H_d^* \tilde{B}_1 \tilde{h}_d - \frac{g_2}{\sqrt{2}} H_d^* \tilde{W}_1^a \sigma^a \tilde{h}_d \nn\\
& - \lambda_S H_u \cdot \tilde{B}_2 \tilde{h}_d - \lambda_S \tilde{h}_u \cdot \tilde{B}_2 H_d - \lambda_T H_d \cdot \tilde{W}_2^a \sigma^a \tilde{h}_u -  \lambda_T \tilde{h}_d \cdot \tilde{W}_2^a \sigma^a H_u + h.c.
\end{align}
We choose to make the definition 
\begin{align}
\Phi_2 = H_u, \qquad \Phi_1^i = -\epsilon_{ij} (H_d^j)^* \leftrightarrow \twv[H_d^0,H_d^-] = \twv[\Phi_1^0,-(\Phi_1^+)^*] 
\end{align}
meaning $H_u \cdot H_d \leftrightarrow - \Phi_1^\dagger \Phi_2 $, which leads to the identifications
\begin{align}
g_{1d}^{11} = 0, \qquad g_{1d}^{21} = \sqrt{2} \lambda_S, \qquad g_{1u}^{11} = - \sqrt{2} \lambda_S, \qquad g_{1u}^{21} = 0 \label{EQ:MatchingGs1} \\
g_{1d}^{12} = g_Y, \qquad g_{1d}^{22} =0, \qquad g_{1u}^{12} = 0, \qquad g_{1u}^{22} = g_Y \label{EQ:MatchingGs2}\\
g_{2d}^{11} = 0, \qquad g_{2d}^{21} = \sqrt{2} \lambda_T, \qquad g_{2u}^{11} = \sqrt{2} \lambda_T, \qquad g_{2u}^{21} =0 \label{EQ:MatchingGs3}\\
g_{2d}^{12} = -g_2, \qquad g_{2d}^{22} = 0, \qquad g_{2u}^{12} = 0, \qquad g_{2u}^{22} =g_2.\label{EQ:MatchingGs4}
\end{align}
These are, however, given in terms of the \DR parameters: making the conversion to \MS we find
\begin{align}
(\tilde{g}_{1u,d}^{ij})_{\MS} =& (\tilde{g}_{1u,d}^{ij})_{\DR} \bigg[ 1 - \frac{1}{4}\frac{g^2_Y}{32\pi^2} - \frac{3}{4} \frac{g^2_2}{32\pi^2} \bigg] \nn\\
(\tilde{g}_{2u,d}^{ij})_{\MS} =& (\tilde{g}_{2u,d}^{ij})_{\DR} \bigg[ 1 - \frac{1}{4}\frac{g^2_Y}{32\pi^2} + \frac{5}{4} \frac{g^2_2}{32\pi^2} \bigg].
\end{align}

\section{Threshold corrections}
\label{APP:THRESHOLDS}

In this section we give the one-loop threshold corrections to the couplings in our theory. Throughout we use the definitions
\begin{align}
\kappa \equiv& \frac{1}{16\pi^2} \nn\\
\blog\ x \equiv& \log \frac{x}{\mu^2} \nn\\
P_{SS} (x,y) \equiv& \frac{ x\blog x - y \blog y}{x-y} - 1,
\end{align}
where $\mu$ is the renormalisation scale at which the quantities are evaluated. 

\subsection{Conversion from \MS to \DR}
\label{app:MSDR}

The conversion of the gauge couplings from the \MS to \DR renormalisation scheme is given by 
\begin{align}
(g_Y)_{\ov{\rm MS}} =&  (g_Y)_{\ov{\rm DR}} \nn\\
(g_2)_{\ov{\rm MS}} =&  (g_2)_{\ov{\rm DR}} \bigg[ 1 - \frac{\kappa g^2_2}{3}\bigg] \nn\\
(g_3)_{\ov{\rm MS}} =&  (g_3)_{\ov{\rm DR}} \bigg[ 1 - \frac{\kappa g^2_3}{2}\bigg].
\end{align}
For the Yukawa couplings, we retain only the strong gauge coupling dependence:
\begin{align}
y^{t,b}_{\ov{\rm MS}} \simeq & y^{t,b}_{\ov{\rm DR}}\bigg( 1 + \frac{4}{3} \kappa g^2_3  \bigg) .
\end{align}

For the Higgs quartic couplings, we define 
\begin{align}
\lambda_i^{\MS} =& \lambda_i^{\DR} + \delta \lambda_i
\end{align}
and then
\begin{align}
\delta \lambda_1 = \delta \lambda_2 = & -\frac{\kappa}{4} \bigg( g_Y^4 + 3 g_2^4 + g_Y^2 g_2^2\bigg) \nn\\
\delta \lambda_3 =& -\frac{\kappa}{4} \bigg( g_Y^4 + 3 g_2^4 -2g_Y^2 g_2^2\bigg)\nn\\
\delta \lambda_4 =& -\kappa g_Y^2 g_2^2.
\label{EQ:lambdaMSDR}
\end{align}
If we express the quartic couplings in terms of the $\MS$ gauge couplings at tree level, then we have a further shift from the shift to $g_2$ of $+\frac{\kappa}{6} g_2^4 $ for $\lambda_{1,2}$ and $- \frac{\kappa}{6} g_2^4 $ for $\lambda_3$.

\subsection{Squark contributions}

\subsubsection{Matching at the SUSY scale}

In the limit that we take in the body of the paper, all of the threshold corrections coming from squarks vanish at the matching scale. However, to extend the results of \cite{Haber:1993an,Lee:2015uza} to our model, we have computed the corrections coming from stops, sbottoms and staus to the quartic couplings allowing non-zero squark trilinears and $\mu$. They are given by
\begin{align}
\delta \lambda_i \equiv& \delta_{\rm th}^{(1)} \lambda_i + \delta_{\Phi}^{(1)} \lambda_i,
\end{align}
where the $\delta_{\rm th}^{(1)} \lambda_i$ contributions are those from bubble, triangle and box diagrams and are unchanged from the MSSM case given in \cite{Haber:1993an,Lee:2015uza}, while the $ \delta_{\Phi}^{(1)} \lambda_i$ are the wavefunction corrections that \emph{are} modified for our model:
\begin{align}
\kappa^{-1} \delta_\Phi^{(1)} \lambda_1 =& - \frac{g_2^2 + g_Y^2}{12M_S^2} ( 3A_b^2 + 3y_t^2\mu^2 + A_\tau^2) \nn\\
\kappa^{-1} \delta_\Phi^{(1)} \lambda_2 =& - \frac{g_2^2 + g_Y^2}{12M_S^2} ( 3A_t^2 + 3y_b^2\mu^2 + \mu^2 y_\tau^2) \nn\\
\kappa^{-1} \delta_\Phi^{(1)} \lambda_3 =& - \frac{g_2^2 - g_Y^2 + 8 \lambda_T^2}{24M_S^2} ( 3A_t^2 +3 A_b^2 + 3(y_b^2 + y_t^2) \mu^2 + A_\tau^2 +y_\tau^2 \mu^2) \nn\\
\kappa^{-1} \delta_\Phi^{(1)} \lambda_4 =& \frac{g_2^2 -2 \lambda_S^2 + 2 \lambda_T^2}{12M_S^2} ( 3A_t^2 +3 A_b^2 + 3(y_b^2 + y_t^2) \mu^2 + A_\tau^2 + y_\tau^2 \mu^2) \nn\\
\kappa^{-1} \delta_\Phi^{(1)} \lambda_5 =& 0 \nn\\
\kappa^{-1} \delta_\Phi^{(1)} \lambda_6 =& \frac{\lambda_S^2 + \lambda_T^2}{12M_S^2} \mu ( 3A_t y_t + 3A_b y_b+ A_\tau y_\tau) \nn\\
\kappa^{-1} \delta_\Phi ^{(1)}\lambda_7 =& \kappa^{-1} \delta_\Phi^{(1)} \lambda_6.
\end{align}

\subsubsection{Matching at a general scale}

If the squarks are not degenerate or we integrate them out at a scale other than a common SUSY scale, then in our limit we have
\begin{align}
\kappa^{-1} \delta \lambda_1 =& \frac{1}{24} \bigg[ ( 9 g_2^4 + g_Y^4 - 36 g_2^2 y_b^2 - 12 g_Y^2 y_b^2 + 72 y_b^4) \blog m_{Q}^2 + 8 g_Y^4 \blog m_{U}^2 + 2 (g_Y^2 - 6 y_b^2)^2 \blog m_{D}^2 \bigg] \nn\\
& + \frac{1}{8} \bigg[ 2 (g_Y^2 - 2 y_\tau^2)^2 \blog m_E^2 + ( g_2^4 + g_Y^4  - 4 g_2^2 y_\tau^2 + 4 g_Y^2 y_\tau^2 + 8 y_\tau^4) \blog m_L^2 \bigg]\nn\\
\kappa^{-1} \delta \lambda_2 =& \frac{1}{24} \bigg[ ( 9 g_2^4 + g_Y^4 - 36 g_2^2 y_t^2 - 12 g_Y^2 y_t^2 + 72 y_t^4) \blog m_{Q}^2 + 2 g_Y^4 \blog m_{D}^2 + 8 (g_Y^2 - 3 y_t^2)^2 \blog m_{U}^2 \bigg]\nn\\
& + \frac{1}{8} \bigg[ 2 g_Y^4 \blog m_E^2 + ( g_2^4 + g_Y^4  ) \blog m_L^2 \bigg]\nn\\
\kappa^{-1} \delta \lambda_3 =& 3 y_b^2 y_t^2 P_{SS} (m_{D}^2, m_{U}^2) \nn\\
&+ \frac{1}{24} \bigg[ ( 9 g_2^4 - g_Y^4 + 72 y_t^2 y_b^2 + 6 g_Y^2 (y_b^2 - y_t^2) -18 g_2^2 (y_b^2 + y_t^2)) \blog m_{Q}^2  \nn\\
&-2g_Y^2 (g_Y^2 - 6 y_b^2) \blog m_{D}^2 - 8 g_Y^2 (g_Y^2 - 3 y_t^2) \blog m_{U}^2 \bigg] \nn\\
& + \frac{1}{8} \bigg[ -2 g_Y^2 (g_Y^2 - 2 y_\tau^2) \blog m_E^2 + (g_2^2 + g_Y^2)( g_2^2 -  g_Y^2 - 2 y_\tau^2 ) \blog m_L^2 \bigg]\nn\\
\kappa^{-1} \delta \lambda_4 =& - 3 y_b^2 y_t^2 P_{SS} (m_{D}^2, m_{U}^2)- \frac{3}{4} \bigg[ (g_2^2 - 2 y_b^2)(g_2^2 - 2 y_t^2)\bigg]\blog m_{Q}^2  \nn\\
& - \frac{1}{4} g_2^2 (g_2^2 - 2 y_\tau^2) \blog m_L^2.
\end{align}
If we consider all squarks to be at a common SUSY scale $M_S$, then these simplify to
\begin{align}
\kappa^{-1} \delta \lambda_1 =& \frac{1}{6} \blog M_S^2 \bigg[ 3 g_2^4 + 5 g_Y^4 - 3 (g_2^2+ g_Y^2) (3y_b^2 + y_\tau^2) + 12( 3 y_b^4 + y_\tau^4) \bigg]\nn\\
 \kappa^{-1} \delta \lambda_2 =& \frac{1}{6} \blog M_S^2 \bigg[ 3 g_2^4 + 5 g_Y^4 - 9 g_2^2 y_t^2 -15 g_Y^2 y_t^2 + 36 y_t^2\bigg]\nn\\
 \kappa^{-1} \delta \lambda_3 =& \frac{1}{12} \blog M_S^2 \bigg[ 6 g_2^4 - 10 g_Y^4 + 72 y_b^2 y_t^2 - 3 (g_2^2 + g_Y^2) (3 y_b^2 + 3y_t^2 + y_\tau^2)) \bigg] \nn\\
 \kappa^{-1} \delta \lambda_4 =&  \frac{1}{2} \blog M_S^2 \bigg[ -2 g_2^4  -12  y_b^2 y_t^2 + g_2^2 (3 y_b^2 + 3y_t^2 + y_\tau^2)) \bigg] \nn\\
 \kappa^{-1} (\delta \lambda_3 + \delta \lambda_4) =& \frac{1}{12} \blog M_S^2 \bigg[ -6 g_2^4 - 10 g_Y^4 + 3 (g_2^2 - g_Y^2) (3 y_b^2 + 3y_t^2 + y_\tau^2)) \bigg].
\end{align}

\subsection{Contributions from the $S,T$ scalars}
\label{app:ST}

Here we present the contributions to the quartic couplings coming from the adjoint scalars $S,T$, in the limit $m_{DY}, m_{D2} \ll m_S, m_T, B_S, B_T$ and assuming no CP-violation. The scalars have masses 
\begin{align}
m_{SR}^2 =& m_S^2 + B_S + 4 m_{DY}^2 \simeq m_S^2 + B_S, \qquad m_{SI}^2 = m_S^2 - B_S \\
m_{TP}^2 =& m_T^2 + B_T + 4 m_{D2}^2 \simeq m_T^2 + B_T, \qquad m_{TM}^2 = m_T^2 - B_T. 
\end{align}

The loop corrections to the quartic couplings are:
\begin{align}
\delta \lambda_1 = \delta  \lambda_2 =\frac{1}{16\pi^2} \frac{1}{2} \bigg[& \lambda_S^4 \log \frac{m_{SR}^2 m_{SI}^2}{\mu^4} + 3 \lambda_T^4 \log \frac{m_{TP}^2 m_{TM}^2}{\mu^4} + (g_2^2 - 2 \lambda_T^2)^2 P_{SS} (m_{TM}^2, m_{TP}^2) \nn\\
& + 2 \lambda_{S}^2 \lambda_T^2 \bigg( P_{SS} (m_{SR}^2, m_{TP}^2) + P_{SS} (m_{SI}^2, m_{TM}^2) \bigg) \bigg] \nn\\
\delta \lambda_3 = \frac{1}{16\pi^2} \frac{1}{2} \bigg[& \lambda_S^4 \log \frac{m_{SR}^2 m_{SI}^2}{\mu^4} + 3 \lambda_T^4 \log \frac{m_{TP}^2 m_{TM}^2}{\mu^4} + (g_2^2 - 2 \lambda_T^2)^2 P_{SS} (m_{TM}^2, m_{TP}^2) \nn\\
& - 2 \lambda_{S}^2 \lambda_T^2 \bigg( P_{SS} (m_{SR}^2, m_{TP}^2) + P_{SS} (m_{SI}^2, m_{TM}^2) \bigg) \bigg] \nn\\
\delta \lambda_4 = \frac{1}{16\pi^2} \bigg[& - (g_2^2 - 2 \lambda_T^2)^2  P_{SS} (m_{TM}^2, m_{TP}^2) \nn\\
& + 2 \lambda_{S}^2 \lambda_T^2 \bigg( P_{SS} (m_{SR}^2, m_{TP}^2) + P_{SS} (m_{SI}^2, m_{TM}^2) \bigg) \bigg]. 
\end{align}
These results update those previously given in the literature by including the electroweak contributions.

\section{One-loop RGEs}

For our numerical study we use two-loop RGEs throughout, as generated by \SARAH. They are too long to put into print; however, for illustration we provide here the one-loop expressions for the low-energy theory of the THDM with electroweakinos, after making the simplification that:
\begin{itemize}
\item Only third generation Yukawa couplings are included.
\item No CP-violation, hence all couplings real.
\item Once we respect the matching conditions (\ref{EQ:MatchingGs1}-\ref{EQ:MatchingGs4}), the beta functions for the couplings that are zero at the supersmmetry scale are zero along the flow, and hence we set the couplings $g_{1d}^{11}$, $g_{1u}^{21}$, $g_{1d}^{22}$, $g_{1u}^{12}$, $g_{2d}^{11}$, $g_{2u}^{21}$, $g_{2d}^{22}$, $g_{2u}^{12} $ to zero.
\end{itemize}
We then define
\begin{align}
\frac{d x}{d\log \mu} \equiv \kappa \beta_x^{(1)} + \kappa^2 \beta_x^{(2)} + ...
\end{align}
and give below the RGEs for the dimensionless quantities in the theory.

\subsection{Gauge couplings}
\begin{flalign}
&\begin{aligned}
\begin{split}
\qquad \qquad \beta_{g_Y}^{(1)} &=  
\frac{23}{3} g_Y^{3} 
\end{split}\\[2mm]
\begin{split}
\qquad \qquad \beta_{g_2}^{(1)} &=  
\frac{1}{3} g_{2}^{3}
\end{split}\\[2mm]
\begin{split}
\qquad \qquad \beta_{g_3}^{(1)} &=  
-7 g_{3}^{3}
\end{split}
\end{aligned}&&
\end{flalign}

\subsection{Yukawa couplings}

\begin{align}
\beta^{(1)}_{y_b} =& \frac{1}{12} y_b (-27 g_2^2 -96 g_3^2 -5 g_Y^2 +54 y_b^2 +6 y_t^2 +12 y_\tau^2 +6 (g_{1d}^{12})^2 +6 (g_{1u}^{11})^2 +18 (g_{2d}^{12})^2 +18 (g_{2u}^{11})^2 )\\ 
\beta^{(1)}_{y_t} =& \frac{1}{12} y_t (-27 g_2^2 -96 g_3^2 -17 g_Y^2 +6 y_b^2 +54 y_t^2 +6 (g_{1d}^{21})^2 +6 (g_{1u}^{22})^2 +18 (g_{2d}^{21})^2 +18 (g_{2u}^{22})^2 )\nn \\ 
\beta^{(1)}_{y_\tau} =& \frac{1}{4} y_\tau (-9 g_2^2 -15 g_Y^2 +2 (6 y_b^2 +5 y_\tau^2 +(g_{1d}^{12})^2+(g_{1u}^{11})^2+3 (g_{2d}^{12})^2 +3 (g_{2u}^{11})^2 ))\nn \\ 
\beta^{(1)}_{g_{1d}^{12}} =& \frac{1}{20} (10 g_{1u}^{22} (2 g_{1u}^{11} g_{1d}^{21} +g_{1d}^{12} g_{1u}^{22}+6 g_{2u}^{11} g_{2d}^{21} )\nn\\
&+5 g_{1d}^{12} (-9 g_2^2 -3 g_Y^2 +12 y_b^2 +4 y_\tau^2 +5 (g_{1d}^{12})^2 +2 (g_{1u}^{11})^2 +9 (g_{2d}^{12})^2 +6 (g_{2u}^{11})^2 +(g_{1d}^{21})^2+3 (g_{2d}^{21})^2 ))\nn \\ 
\beta^{(1)}_{g_{1u}^{11}} =& \frac{1}{20} (10 g_{1d}^{21} (g_{1u}^{11} g_{1d}^{21} +2 g_{1d}^{12} g_{1u}^{22} +6 g_{2d}^{12} g_{2u}^{22} )\nn\\
&+5 g_{1u}^{11} (-9 g_2^2 -3 g_Y^2 +12 y_b^2 +4 y_\tau^2 +2 (g_{1d}^{12})^2 +5 (g_{1u}^{11})^2 +6 (g_{2d}^{12})^2 +9 (g_{2u}^{11})^2 +(g_{1u}^{22})^2+3 (g_{2u}^{22})^2 ))\nn \\ 
\beta^{(1)}_{g_{2d}^{12}} =& \frac{1}{20} (5 g_{2d}^{12} (-33 g_2^2 -3 g_Y^2 +12 y_b^2 +4 y_\tau^2 +3 (g_{1d}^{12})^2 +2 (g_{1u}^{11})^2 +11 (g_{2d}^{12})^2 +6 (g_{2u}^{11})^2 +(g_{1d}^{21})^2+3 (g_{2d}^{21})^2 )\nn\\
&+10 g_{2u}^{22} (2 g_{1u}^{11} g_{1d}^{21} -2 g_{2u}^{11} g_{2d}^{21} +g_{2d}^{12} g_{2u}^{22} ))\nn \\ 
\beta^{(1)}_{g_{2u}^{11}} =& \frac{1}{20} (10 g_{2d}^{21} (2 g_{1d}^{12} g_{1u}^{22} +g_{2u}^{11} g_{2d}^{21} -2 g_{2d}^{12} g_{2u}^{22} )\nn\\
&+5 g_{2u}^{11} (-33 g_2^2 -3 g_Y^2 +12 y_b^2 +4 y_\tau^2 +2 (g_{1d}^{12})^2 +3 (g_{1u}^{11})^2 +6 (g_{2d}^{12})^2 +11 (g_{2u}^{11})^2 +(g_{1u}^{22})^2+3 (g_{2u}^{22})^2 ))\nn \\ 
\beta^{(1)}_{g_{1d}^{21}} =& \frac{1}{20} (10 g_{1u}^{11} (g_{1u}^{11} g_{1d}^{21} +2 g_{1d}^{12} g_{1u}^{22} +6 g_{2d}^{12} g_{2u}^{22} )\nn\\
&+5 g_{1d}^{21} (-9 g_2^2 -3 g_Y^2 +12 y_t^2 +(g_{1d}^{12})^2+3 (g_{2d}^{12})^2 +5 (g_{1d}^{21})^2 +2 (g_{1u}^{22})^2 +9 (g_{2d}^{21})^2 +6 (g_{2u}^{22})^2 ))\nn \\ 
\beta^{(1)}_{g_{1u}^{22}} =& \frac{1}{4} (2 (g_{1d}^{12})^2 g_{1u}^{22} +4 g_{1d}^{12} (g_{1u}^{11} g_{1d}^{21} +3 g_{2u}^{11} g_{2d}^{21} )\nn\\
&+g_{1u}^{22} (-9 g_2^2 -3 g_Y^2 +12 y_t^2 +(g_{1u}^{11})^2+3 (g_{2u}^{11})^2 +2 (g_{1d}^{21})^2 +5 (g_{1u}^{22})^2 +6 (g_{2d}^{21})^2 +9 (g_{2u}^{22})^2 ))\nn \\ 
\beta^{(1)}_{g_{2d}^{21}} =& \frac{1}{20} (10 g_{2u}^{11} (2 g_{1d}^{12} g_{1u}^{22} +g_{2u}^{11} g_{2d}^{21} -2 g_{2d}^{12} g_{2u}^{22} )\nn\\
&+5 g_{2d}^{21} (-33 g_2^2 -3 g_Y^2 +12 y_t^2 +(g_{1d}^{12})^2+3 (g_{2d}^{12})^2 +3 (g_{1d}^{21})^2 +2 (g_{1u}^{22})^2 +11 (g_{2d}^{21})^2 +6 (g_{2u}^{22})^2 ))\nn \\ 
\beta^{(1)}_{g_{2u}^{22}} =& \frac{1}{4} (4 g_{1u}^{11} g_{2d}^{12} g_{1d}^{21} -4 g_{2d}^{12} g_{2u}^{11} g_{2d}^{21} +(g_{1u}^{11})^2 g_{2u}^{22} +2 (g_{2d}^{12})^2 g_{2u}^{22} \nn\\
&+g_{2u}^{22} (-33 g_2^2 -3 g_Y^2 +12 y_t^2 +3 (g_{2u}^{11})^2 +2 (g_{1d}^{21})^2 +3 (g_{1u}^{22})^2 +6 (g_{2d}^{21})^2 +11 (g_{2u}^{22})^2 )).\nn
\end{align}

\subsection{Quartic scalar couplings}

\begin{align} 
\beta^{(1)}_{\lambda_1} =& \frac{1}{4} \bigg(9 g_2^4 +3 g_Y^4 +6 g_2^2 (g_Y^2-6 \lambda_1 )-12 g_Y^2 \lambda_1\bigg) -\bigg(12 y_b^4 +4 y_\tau^4 -12 y_b^2 \lambda_1 -4 y_\tau^2 \lambda_1 -12 \lambda_1^2 \\
&-4 \lambda_3^2 -4 \lambda_3 \lambda_4 -2 \lambda_4^2 -2 \lambda_5^2 -2 \lambda_1 (g_{1d}^{12})^2 +(g_{1d}^{12})^4-2 \lambda_1 (g_{1u}^{11})^2 +(g_{1u}^{11})^4-6 \lambda_1 (g_{2d}^{12})^2 \nn\\
&+2 (g_{1d}^{12})^2 (g_{2d}^{12})^2 +5 (g_{2d}^{12})^4 -6 \lambda_1 (g_{2u}^{11})^2 +2 (g_{1u}^{11})^2 (g_{2u}^{11})^2 +5 (g_{2u}^{11})^4 \bigg)\nn \\ 
\beta^{(1)}_{\lambda_2} =& \frac{1}{4} \bigg(9 g_2^4 +3 g_Y^4 +6 g_2^2 (g_Y^2-6 \lambda_2 )-12 g_Y^2 \lambda_2 \bigg) - \bigg(12 y_t^4 -12 y_t^2 \lambda_2 -12 \lambda_2^2 -4 \lambda_3^2 -4 \lambda_3 \lambda_4 \nn\\
& -2 \lambda_4^2 -2 \lambda_5^2 +(g_{1d}^{21})^4+(g_{1u}^{22})^4+2 (g_{1d}^{21})^2 (g_{2d}^{21})^2 +5 (g_{2d}^{21})^4 +2 (g_{1u}^{22})^2 (g_{2u}^{22})^2 +5 (g_{2u}^{22})^4 \nn\\
&-2 \lambda_2 \big[(g_{1d}^{21})^2+(g_{1u}^{22})^2+3 ((g_{2d}^{21})^2+(g_{2u}^{22})^2)\big]\bigg)\nn \\ 
\beta^{(1)}_{\lambda_3} =& \frac{9}{4} g_2^4 -\frac{3}{2} g_2^2 g_Y^2 +\frac{3}{4} g_Y^4 -12 y_b^2 y_t^2 -9 g_2^2 \lambda_3 -3 g_Y^2 \lambda_3 +6 y_b^2 \lambda_3 +6 y_t^2 \lambda_3 +2 y_\tau^2 \lambda_3 +6 \lambda_1 \lambda_3 +6 \lambda_2 \lambda_3 +4 \lambda_3^2 \nn\\
&+2 \lambda_1 \lambda_4 +2 \lambda_2 \lambda_4 +2 \lambda_4^2 +2 \lambda_5^2 +3 \lambda_3 (g_{2d}^{12})^2 +3 \lambda_3 (g_{2u}^{11})^2 +\lambda_3 (g_{1d}^{21})^2 - (g_{1u}^{11})^2 (g_{1d}^{21})^2 -2 (g_{2d}^{12})^2 (g_{1d}^{21})^2\nn\\
& +\lambda_3 (g_{1u}^{22})^2 - (g_{1d}^{12})^2 (g_{1u}^{22})^2 -2 (g_{2u}^{11})^2 (g_{1u}^{22})^2 +2 g_{1u}^{11} g_{2u}^{11} g_{1d}^{21} g_{2d}^{21} -4 g_{1d}^{12} g_{2u}^{11} g_{1u}^{22} g_{2d}^{21} +3 \lambda_3 (g_{2d}^{21})^2 -4 (g_{2d}^{12})^2 (g_{2d}^{21})^2 \nn\\
&-5 (g_{2u}^{11})^2 (g_{2d}^{21})^2  +(g_{1d}^{12})^2 (\lambda_3-2 (g_{2d}^{21})^2 )-4 g_{1u}^{11} g_{2d}^{12} g_{1d}^{21} g_{2u}^{22} +2 g_{1d}^{12} g_{2d}^{12} g_{1u}^{22} g_{2u}^{22} +8 g_{2d}^{12} g_{2u}^{11} g_{2d}^{21} g_{2u}^{22}\nn\\
& +3 \lambda_3 (g_{2u}^{22})^2 -5 (g_{2d}^{12})^2 (g_{2u}^{22})^2 -4 (g_{2u}^{11})^2 (g_{2u}^{22})^2 +(g_{1u}^{11})^2 (\lambda_3-2 (g_{2u}^{22})^2 )\nn \\ 
\beta^{(1)}_{\lambda_4} =& 3 g_2^2 g_Y^2 +12 y_b^2 y_t^2 -9 g_2^2 \lambda_4 -3 g_Y^2 \lambda_4 +6 y_b^2 \lambda_4 +6 y_t^2 \lambda_4 +2 y_\tau^2 \lambda_4 +2 \lambda_1 \lambda_4 +2 \lambda_2 \lambda_4 +8 \lambda_3 \lambda_4 +4 \lambda_4^2 +8 \lambda_5^2\nn\\
& +3 \lambda_4 (g_{2d}^{12})^2 +3 \lambda_4 (g_{2u}^{11})^2 +\lambda_4 (g_{1d}^{21})^2 +(g_{2d}^{12})^2 (g_{1d}^{21})^2 -2 g_{1d}^{12} g_{1u}^{11} g_{1d}^{21} g_{1u}^{22} +\lambda_4 (g_{1u}^{22})^2 +(g_{2u}^{11})^2 (g_{1u}^{22})^2 \nn\\
&-4 g_{1u}^{11} g_{2u}^{11} g_{1d}^{21} g_{2d}^{21} +2 g_{1d}^{12} g_{2u}^{11} g_{1u}^{22} g_{2d}^{21} +3 \lambda_4 (g_{2d}^{21})^2 - (g_{2d}^{12})^2 (g_{2d}^{21})^2 +4 (g_{2u}^{11})^2 (g_{2d}^{21})^2 \nn\\
&+(g_{1d}^{12})^2 (\lambda_4- (g_{1d}^{21})^2 +(g_{2d}^{21})^2)+2 g_{1u}^{11} g_{2d}^{12} g_{1d}^{21} g_{2u}^{22} -4 g_{1d}^{12} g_{2d}^{12} g_{1u}^{22} g_{2u}^{22} -10 g_{2d}^{12} g_{2u}^{11} g_{2d}^{21} g_{2u}^{22}\nn\\
& +3 \lambda_4 (g_{2u}^{22})^2 +4 (g_{2d}^{12})^2 (g_{2u}^{22})^2 - (g_{2u}^{11})^2 (g_{2u}^{22})^2 +(g_{1u}^{11})^2 (\lambda_4- (g_{1u}^{22})^2 +(g_{2u}^{22})^2)\nn \\ 
\beta^{(1)}_{\lambda_5} =& \lambda_5 \bigg(-9 g_2^2 -3 g_Y^2 +6 y_b^2 +6 y_t^2 +2 y_\tau^2 +2 \lambda_1 +2 \lambda_2 +8 \lambda_3 +12 \lambda_4 \nn\\
&+(g_{1d}^{12})^2+(g_{1u}^{11})^2+3 (g_{2d}^{12})^2 +3 (g_{2u}^{11})^2 +(g_{1d}^{21})^2+(g_{1u}^{22})^2+3 (g_{2d}^{21})^2 +3 (g_{2u}^{22})^2 \bigg). \nn
\end{align}

\section{MRSSM corrections}
\label{APP:MRSSM}

Here we collect the tree-level and leading one-loop threshold corrections to the THDM paramters in the MRSSM.

\subsection{Tree-level}
\label{APP:MRSSM_TreeLevel}

The tree-level $\lambda_i$ are given by
\begin{align}
\lambda_1 = \lambda_2 = \frac{1}{4} ( g_2^2 + g_Y^2), \qquad \lambda_3 = \frac{1}{4} ( g_2^2 - g_Y^2), \qquad \lambda_4 = - \frac{1}{2} g_2^2.
\end{align}
The shifts from integrating out the adjoint scalars give
\begin{align}
\delta \lambda_1 =& - \frac{ (g_Y m_{DY} - \sqrt{2} \lambda_{S_d} \mu_d)^2}{m_{SR}^2} - \frac{ (g_2 m_{D2} + \sqrt{2} \lambda_{T_d} \mu_d)^2}{m_{TP}^2} \nn\\
\delta \lambda_2 =& - \frac{ (g_Y m_{DY} + \sqrt{2} \lambda_{S_u} \mu_u)^2}{m_{SR}^2} - \frac{ (g_2 m_{D2} + \sqrt{2} \lambda_{T_u} \mu_u)^2}{m_{TP}^2} \nn\\
\delta \lambda_3 =& - \frac{A_{T}^2}{m_{TM}^2} - \frac{A_{T}^2}{m_{TP}^2} \nn\\
&+ 
\frac{ (g_Y m_{DY} - \sqrt{2} \lambda_{S_d} \mu_d)(g_Y m_{DY} + \sqrt{2} \lambda_{S_u} \mu_u)}{m_{SR}^2}- \frac{ (g_2 m_{D2} + \sqrt{2} \lambda_{T_d} \mu_d)(g_2 m_{D2} + \sqrt{2} \lambda_{T_u} \mu_u)}{m_{TP}^2} \nn\\
\delta \lambda_4 =& -\frac{A_{S}^2}{2 m_{SI}^2} - \frac{A_{S}^2}{2 m_{SR}^2} + \frac{A_{T}^2}{2 m_{TM}^2} + \frac{A_{T}^2}{2 m_{TP}^2}  \nn\\
&+ 2\frac{ (g_2 m_{D2} + \sqrt{2} \lambda_{T_d} \mu_d)(g_2 m_{D2} + \sqrt{2} \lambda_{T_u} \mu_u)}{m_{TP}^2} \nn\\
\delta \lambda_5 =& \frac{A_{S}^2}{2 m_{SI}^2} - \frac{A_{S}^2}{2 m_{SR}^2} +\frac{A_{T}^2}{2 m_{TM}^2} - \frac{A_{T}^2}{2 m_{TP}^2} \nn\\
\delta \lambda_6 =& \frac{A_{S} ( - g_Y m_{DY} + \sqrt{2} \lambda_{S_d} \mu_d)}{\sqrt{2} m_{SR}^2} + \frac{A_{T} ( g_2 m_{D2} + \sqrt{2} \lambda_{T_d} \mu_d)}{\sqrt{2} m_{TP}^2} \nn\\
\delta \lambda_7 =& \frac{A_{S} ( g_Y m_{DY} + \sqrt{2} \lambda_{S_u} \mu_u)}{\sqrt{2} m_{SR}^2} - \frac{A_{T} ( g_2 m_{D2} + \sqrt{2} \lambda_{T_u} \mu_u )}{\sqrt{2} m_{TP}^2}
\end{align}

\subsection{One-loop}
\label{APP:MRSSM_OneLoop}

The one-loop corrections from the adjoint scalars in the limit that we can neglect the Dirac gaugino masses are given by:
\begin{align}
\kappa^{-1} \delta \lambda_1 =& \frac{1}{2} \bigg[ 3 \lambda_{T_d}^4 \log \frac{(m_{TM}^2 m_{TP}^2)}{\mu^4}  + \lambda_{S_d}^4 \log \frac{(m_{SR}^2 m_{SI}^2)}{\mu^4} + 2 \lambda_{S_d}^2 \lambda_{T_d}^2 \bigg( P_{SS} (m_{SR}^2,m_{TP}^2 ) + P_{SS} (m_{SI}^2,m_{TM}^2 )\bigg)\nn\\
& + (g_2^2 - 2 \lambda_{T_d}^2)^2  P_{SS} (m_{TM}^2 ,m_{TP}^2) \bigg]\nn\\
\kappa^{-1}  \delta\lambda_2 =& \frac{1}{2} \bigg[ 3 \lambda_{T_u}^4 \log \frac{(m_{TM}^2 m_{TP}^2)}{\mu^4}  + \lambda_{S_u}^4 \log \frac{(m_{SR}^2 m_{SI}^2)}{\mu^4} + 2 \lambda_{S_u}^2 \lambda_{T_u}^2 \bigg( P_{SS} (m_{SR}^2,m_{TP}^2 ) + P_{SS} (m_{SI}^2,m_{TM}^2 )\bigg)\nn\\
& + (g_2^2 - 2 \lambda_{T_u}^2)^2  P_{SS} (m_{TM}^2 ,m_{TP}^2) \bigg]\nn\\
\kappa^{-1}  \delta\lambda_3 =& \frac{1}{2} \bigg[ 3 \lambda_{T_u}^2 \lambda_{T_d}^2 \log \frac{(m_{TM}^2 m_{TP}^2)}{\mu^4}  + \lambda_{S_u}^2 \lambda_{S_d}^2  \log \frac{(m_{SR}^2 m_{SI}^2)}{\mu^4}\nn\\
& + 2 \lambda_{S_u} \lambda_{T_u} \lambda_{S_d} \lambda_{T_d} \bigg( P_{SS} (m_{SR}^2,m_{TP}^2 ) + P_{SS} (m_{SI}^2,m_{TP}^2 )\bigg)\nn\\
& + (g_2^2 - 2 \lambda_{T_u}^2)(g_2^2 - 2 \lambda_{T_d}^2)   P_{SS} (m_{TM}^2 ,m_{TP}^2) \bigg]\nn\\
\kappa^{-1}  \delta\lambda_4 =&- \bigg[ 2 \lambda_{S_u} \lambda_{T_u} \lambda_{S_d} \lambda_{T_d} \bigg( P_{SS} (m_{SR}^2,m_{TP}^2 ) + P_{SS} (m_{SI}^2,m_{TM}^2 )\bigg)\nn\\
& + (g_2^2 - 2 \lambda_{T_u}^2)(g_2^2 - 2 \lambda_{T_d}^2)   P_{SS} (m_{TM}^2 ,m_{TP}^2) \bigg]
\end{align}

\bibliographystyle{utphys}
\bibliography{biblio}

\end{document}